\documentclass[%
 reprint,
superscriptaddress,
 amsmath,amssymb,
 aps,pra
]{revtex4-2}
\usepackage{xcolor}
\usepackage{graphicx}
\usepackage{xspace}
\usepackage{xpatch}
\usepackage{tikz}
\usepackage{hyperref}

\usepackage{color}
\usepackage[utf8]{inputenc}
\usepackage{physics}
\usepackage{url}
\usepackage{braket}

\def\Hbar  {\ensuremath{\overline {\mathrm{H}}}\xspace}
\def\gbar  {\ensuremath{\overline {g}}\xspace}

\def\mmax  {\ensuremath{\mathrm{max}}\xspace}
\def\md    {\ensuremath{\mathrm{d}}\xspace}
\def\mr    {\ensuremath{\mathrm{r}}\xspace}
\def\Ai    {\ensuremath{\mathrm{Ai}}\xspace}
\def\MC    {\ensuremath{\mathrm{MC}}\xspace}
\def\CR    {\ensuremath{\mathrm{CR}}\xspace}

\def\br    {\ensuremath{\mathbf{r}}\xspace}

\def\bv    {\ensuremath{\mathbf{v}}\xspace}
\def\bp    {\ensuremath{\mathbf{p}}\xspace}

\def\bq    {\ensuremath{\mathbf{q}}\xspace}
\def\hbq   {\ensuremath{\hat{\mathbf{q}}}\xspace}
\def\hbn   {\ensuremath{\hat{\mathbf{n}}}\xspace}

\def\onv   {\ensuremath{\overline{\overline{v}}}\xspace}
\def\onR   {\ensuremath{\overline{\overline{R}}}\xspace}
\def\obR   {\ensuremath{\overline{\overline{\mathbf{R}}}}\xspace}
\def\obr   {\ensuremath{\overline{\overline{\mathbf{r}}}}\xspace}
\def\obp   {\ensuremath{\overline{\overline{\mathbf{p}}}}\xspace}
\def\obq   {\ensuremath{\overline{\overline{\mathbf{q}}}}\xspace}
\def\obv   {\ensuremath{\overline{\overline{\mathbf{v}}}}\xspace}

\begin{document}
\title{Quantum interference measurement of the free fall of anti-hydrogen}

\author{Olivier Rousselle}
\affiliation{%
 Laboratoire Kastler Brossel, Sorbonne Universit\'e, CNRS, ENS-PSL, Coll\`ege de France, 4 place Jussieu, 75005 Paris, France
 }%
\author{Pierre Clad\'e}%
\affiliation{%
 Laboratoire Kastler Brossel, Sorbonne Universit\'e, CNRS, ENS-PSL, Coll\`ege de France, 4 place Jussieu, 75005 Paris, France 
}%
\author{Sa\"ida Guellati-Kh\'elifa}%
\affiliation{%
 Laboratoire Kastler Brossel, Sorbonne Universit\'e, CNRS, ENS-PSL, Coll\`ege de France, 4 place Jussieu, 75005 Paris, France 
}\affiliation{Conservatoire National des Arts et M\'etiers, 292 rue Saint Martin, 75003 Paris, France
}%
\author{Romain Gu\'erout}
\affiliation{%
 Laboratoire Kastler Brossel, Sorbonne Universit\'e, CNRS, ENS-PSL, Coll\`ege de France, 4 place Jussieu, 75005 Paris, France 
}%
\author{Serge Reynaud}
\affiliation{%
 Laboratoire Kastler Brossel, Sorbonne Universit\'e, CNRS, ENS-PSL, Coll\`ege de France, 4 place Jussieu, 75005 Paris, France 
}%


\begin{abstract}
We analyze a quantum measurement designed to improve the accuracy for the free-fall acceleration of anti-hydrogen in the GBAR experiment. 
Including the effect of photo-detachment recoil in the analysis and developing a full quantum analysis of anti-matter wave propagation, we show that the accuracy is improved by approximately three orders of magnitude with respect to the classical timing technique planned for the current experiment. 
\end{abstract}

\maketitle

\section{Introduction}
\label{sec:introduction}

One of the most important questions of fundamental physics is the asymmetry between matter and antimatter observed in the Universe \cite{Hori2013,Bertsche2015,Charlton2017,Yamazaki2020}. 
In this context, it is extremely important to compare the gravitational properties of antimatter with those of matter \cite{Bondi1957,Morrison1958,Scherk1979,Nieto1991,Adelberger1991,%
Huber2000,Chardin2018}. 
The aim of measuring the free fall acceleration $\gbar$ of anti-hydrogen $\Hbar$ in Earth’s gravitational field has been approached in the last decades \cite{Alpha2013} and indirect indications of the same sign as for matter obtained recently \cite{Borchert2022}. 

Improving the accuracy of the measurement of $\gbar$ will remain a crucial objective for advanced tests of the Equivalence Principle involving antimatter besides matter test masses \cite{Wagner2012,Touboul2017,Will2018,Viswanathan2018}. 
Ambitious projects are currently developed at new CERN facilities to produce low energy anti-hydrogen atoms \cite{Maury2014} and to improve the accuracy of $\gbar$-measurement \cite{Indelicato2014,Bertsche2018,Pagano2020}. Among these projects, the experiment \emph{Gravitational Behaviour of Anti-hydrogen at Rest} (GBAR) aims at timing the free fall of ultra-cold $\Hbar$ atoms, with a precision goal of 1\% \cite{Perez2015}.

The principle of GBAR experiment is based upon an original idea of Walz and Hänsch \cite{Walz2004}. 
Positive ions $\Hbar^+$ are cooled in an ion trap to a low temperature \cite{Hilico2014,Sillitoe2017} and the excess positron is photo-detached by a laser pulse \cite{Lykke1991,Vandevraye2014,Bresteau2017}. This pulse marks the start of the free fall of the $\Hbar$ neutral atom. The free fall on a given height is timed with a stop signal associated with the annihilation of anti-hydrogen reaching the surface of the detector. 
The positions in time and space of the annihilation event are reconstructed from the analysis of the impact of secondary particles on Micromegas and scintillation detectors surrounding the experiment chamber \cite{Radics2019}. 

The precision goal in this timing measurement of classical free fall is mainly limited by the low temperature of the $\Hbar^+$ ions in the ion trap. Recent studies have been devoted to the analysis of accuracy to be expected in this experiment when taking into account the atomic recoil associated to the photo-detachment process \cite{Rousselle2022NJP,Rousselle2022PRA}. These studies have confirmed that the precision goal of 1\% was attainable. 

The current design of the GBAR experiment relies on a classical free fall timing but it has also been proven that free fall acceleration can be measured by matter-wave interferometry  \cite{Kasevich1991,Borde2002,Merlet2010,Asenbaum2020}. 
In this article, we study a quantum interference experiment proposed to improve the accuracy of the GBAR measurement by approximately 3 orders of magnitude with respect to the classical timing measurement \cite{Crepin2019}. The idea is to use quantum techniques such to those utilized on \emph{Whispering Gallery Modes} \cite{Nesvizhevsky2009} and \emph{Gravitational Quantum States} (GQS) of ultra-cold neutrons 
\cite{Nesvizhevsky2002,Nesvizhevsky2003,Nesvizhevsky2005}. GQS are expected to exist for anti-hydrogen atoms, thanks to quantum reflection on the Casimir-Polder potential arising in the vicinity of a material surface \cite{Froelich2012,Dufour2013}.

Atoms with a low vertical velocity above a quantum reflecting mirror placed at a small distance below the $\Hbar$ source will be trapped by the combined action of quantum reflection and gravity \cite{Jurisch2006,Madronero2007}. The quantum paths corresponding to different quantum states above this mirror produce an interference pattern at the end of the mirror. This pattern is revealed by a free fall period, from the end of the mirror to the detection plate placed at a macroscopic distance below the quantum reflection mirror. 
The aims of the present paper are to account for the atomic recoil induced by the photo-detachment process and also to give a full quantum analysis of anti-matter wave propagation, which was treated in \cite{Crepin2019} by a far-field diffraction approximation. 

We first remind in the next section (\S \ref{sec:reminders}) some results which will be needed for our analysis, namely the description of the initial distribution of atomic velocities accounting for the recoil due to the photo-detachment process and the quantum design proposed for the GBAR experiment to improve the accuracy of the classical experiment. We study the pattern appearing at the end of the quantum mirror as a consequence of interference between the different quantum states (\S \ref{sec:interference}) and show that these interferences are still present with photo-detachment recoil. We then deduce the interference pattern seen in the distribution of annihilation events on the detection plate (\S \ref{sec:patterndetector}) by using the full quantum propagator rather than a far-field approximation. We present the estimation of the uncertainty obtained with numerical and analytical statistical methods (\S \ref{sec:uncertainty}).
We finally compare the results of the present paper to those of the classical analysis \cite{Rousselle2022NJP} as well as those of the approximate quantum analysis \cite{Crepin2019}.

\section{Reminders}
\label{sec:reminders}

We first present in this section a few reminders which will be used in the subsequent analysis, namely the description of the initial distribution of atomic velocities and the quantum design for the GBAR experiment. 

\subsection{Description of the initial state}

The description of the initial state requires a careful attention as it mixes a coherent preparation of ions in the ion trap and an incoherent sum over the recoil due to the momentum brought by the photo-detached positron which is not detected in the experiment. This incoherent part of the preparation of the $\Hbar$ atoms could blur the interference pattern predicted without photo-detachment and thus lead to a degradation of the accuracy expected for the quantum measurement. 

The initial state of $\Hbar$ atoms has to be described by a density matrix corresponding to a statistical mixture of different recoils of the coherent state in the ion trap affected by photo-detachment recoil
\begin{equation}
\rho_0 = \int \,\varpi(\hbq) \,\md\Omega 
\ket{\Psi_{\bq,0}}\bra{\Psi_{\bq,0}} ~, 
\label{initiamdensitymatrix}
\end{equation} 
where $\ket{\Psi_{\bq,0}}$ represents the atomic state after the photo-detachment process and the recoil $\bq$, while $\varpi$ is the distribution of this recoil. Both elements are now explained. 

The wave function $\ket{\Psi_{\bq,0}}$ is written in the position representation as a Gaussian centered at $\br_h=(0,0,h)$ as $\Hbar$ atoms are supposed to be released from the trap at height $h=10~\mu$m above the mirror ($\delta\br_0\equiv\br_0-\br_h$)
\begin{equation}
\Psi_{\bq,0}(\br_0)=\left(\frac{1}{2\pi \zeta^2}\right)^{3/4}
\exp\left(-\frac{\delta\br_0^2}{4\zeta^2}
+\frac{\imath}{\hbar}\bq\cdot\delta\br_0\right) ~.
\label{coherentwavepacket}
\end{equation} 
We denote $\zeta=\sqrt{\frac{\hbar}{2m\omega}}$ the position dispersion in the ion trap with $\omega=2\pi f$ and $f$ the trap frequency.

If there were no dispersion of the recoil $\bq$, the density matrix \eqref{initiamdensitymatrix} would represent a perfectly coherent quantum state. Here in contrast, there exists a dispersion of recoil, characterized by the distribution $\varpi$ in \eqref{initiamdensitymatrix}. Details of the photo-detachment process are presented in \cite{Rousselle2022NJP}. 
The recoil $\bq$ transferred to the atom in the photo-detachment process has a fixed magnitude $q$ determined by the excess energy $\delta E$ above the photo-detachment threshold 
\begin{equation}
q=\left\vert\bq\right\vert=\sqrt{2m\delta E}~,
\label{recoilmagnitude}
\end{equation}
where we used the fact that the mass of the positron is much smaller than that of the atom. 
The photo-detachment cross-section scales as the power $\delta E^{3/2}$  of the excess energy \cite{Lykke1991,Vandevraye2014,Bresteau2017} which implies that this parameter has to be large enough. 

The recoil is fixed as $\bq=q~\hbq$ with the magnitude $q$ fixed by Eq.\eqref{recoilmagnitude} and the unit vector aligned on the velocity of the photo-detached positron.
The angular distribution of recoil obeys a dipolar distribution centered around the direction $\hbn$ of the photo-detachment laser polarization (with $\Omega$ the solid angle) 
\begin{equation}
\varpi(\hbq) \,\md\Omega
=3\left(\hbq.\hbn\right)^2  \,\frac{\md\Omega}{4\pi} ~.
\end{equation}

The photo-detachment is considered as a very short process affecting the position of the atom in a negligible way. The momentum distribution is obtained as the convolution product of the Gaussian distribution in the ion trap and of the distribution of photo-detachment recoil. This momentum distribution in studied in details in \cite{Rousselle2022NJP} where an analytical expression is given. From now on, we discuss this distribution in terms of velocity $\displaystyle{\bv_0 = \frac{\bp_0}m}$ instead of momentum. 

\begin{figure}[t!]
\centering
\includegraphics[scale=0.65]{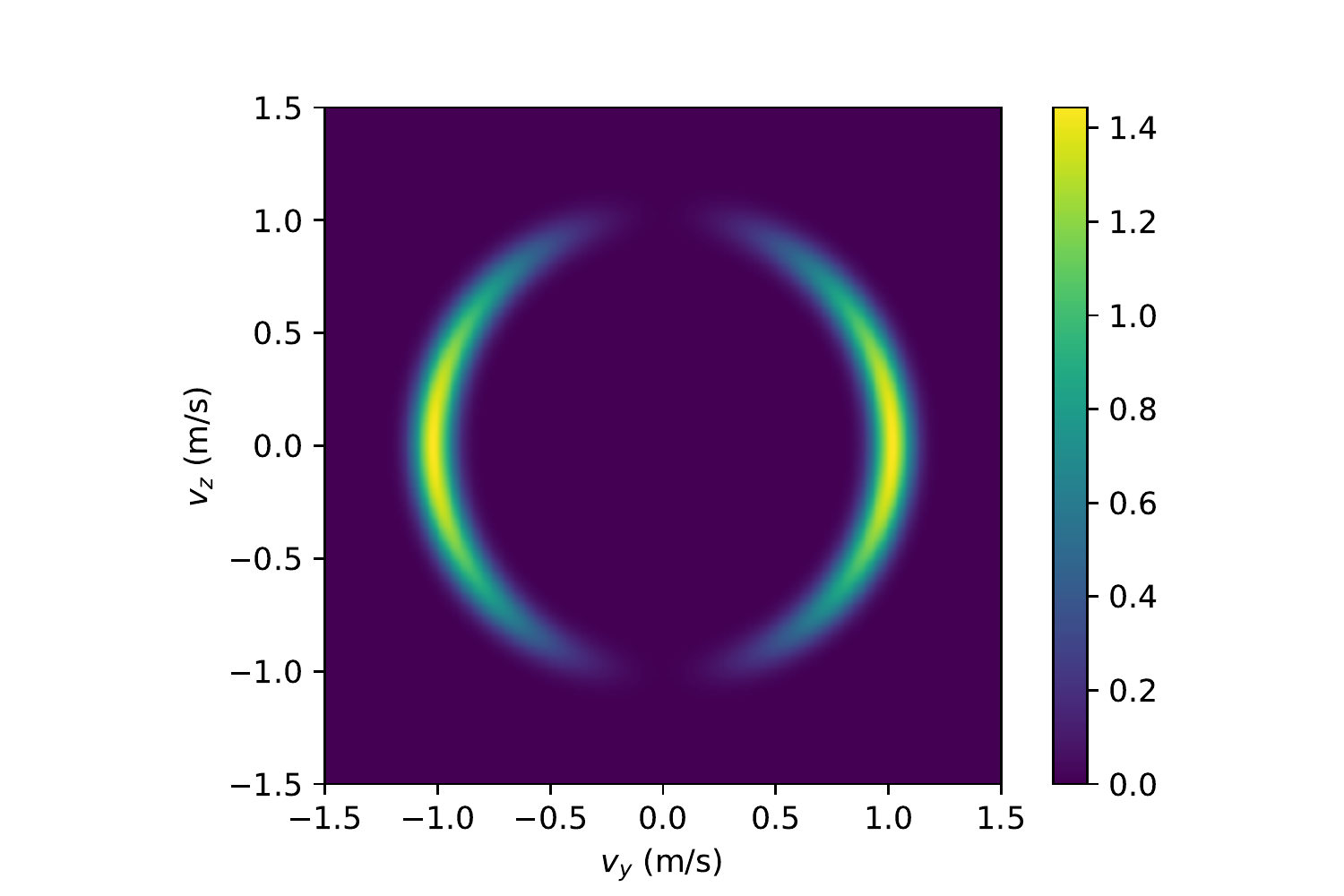}
\caption{Density plot of the velocity distribution $\Pi_0$ in the $\left(v_y,v_z\right)$ plane for $\delta E=10~\mu$eV and $f=20\,$kHz. }
\label{fig:Pi_0-photo-detachment}
\end{figure}

In the following, we use trap frequency $f=20\,$kHz, which corresponds to position dispersion $\zeta\simeq0.5~\mu$m, to conjugated momentum dispersion $\Delta p=\frac\hbar{2\zeta}$ and to velocity dispersion $\Delta v=\frac{\Delta p}{m}\simeq 6.3\,$cm/s. 
We also take $\delta E=10~\mu$eV which corresponds to a recoil velocity $v_\mr =\frac{q}{m}=1.02\,$m/s. 
The velocity distribution $\Pi_0 \left(\bv_0\right)$ is shown as a density plot in Fig.\ref{fig:Pi_0-photo-detachment}. It is a shell with a small width $\Delta v=6.3\,$cm/s around the sphere of radius $v_\mr  = 1.02\,$m/s. Though $v_\mr $ is much larger than $\Delta v$, we will see in the following that the recoil does not degrade too much the precision expected for the measurement of $g$.
We have assumed the laser polarisation to be along the $y-$direction. in order to maximize the proportion of atoms going out of the trap with a nearly horizontal velocity (see discussions in \S\ref{sec:reminders}.B). 

\subsection{Quantum interference design}

The quantum design \cite{Crepin2019} has been chosen to add minimal modifications to the GBAR classical design \cite{Perez2015}. A circular reflecting mirror of diameter $d=5\,$cm is added
at distance $h$ below the center of the ion trap. The experimental setup is then decomposed into two zones: the interference zone above the mirror on which GQS interfere, and the free fall zone with height $H=30\,$cm, which transforms the interference pattern at the end of the mirror into an interference pattern read on positions at the detector.

The quantum design was already studied without accounting for the photo-detachment recoil \cite{Crepin2019}. The overall movement over the quantum mirror was produced by an initial horizontal kick $v_0$. Here, the photo-detachment recoil $v_\mr \simeq1\,$m/s is sufficient to produce the overall movement so that we don't need to consider a further initial kick. For simplicity, we will disregard the kick produced by the photon absorption $v_\gamma=0.24\,$m/s which is small with respect to $v_\mr $. 

\begin{figure}[t!]
\centering
\includegraphics[width=.7\linewidth]{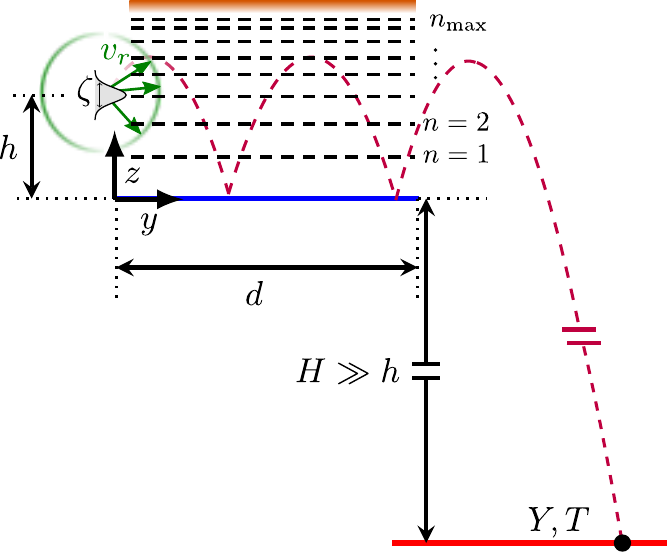}
\caption{Schematic representation of the design. The quantum mirror of radius $d$ is the blue horizontal line. The atoms are released at a mean height $h$ above the mirror with a dispersion $\zeta$ in all space directions. An absorber (orange line) is placed above the mirror allowing $n_\mmax$ Gravitational Quantum States to pass through the device. $H$ is the free-fall height, $Y$ and $T$ the positions in space and time of the detection events on the detector plate in red.}
\label{fig:quantum-design}
\end{figure}

The quantum design is schematized as a 2D view in Fig.\ref{fig:quantum-design}, with lowercase letters representing quantities relative to first stages of preparation and interference above the mirror, while uppercase letters represent quantities associated with free-fall and detection stages. The parabolas in purple represent classical motions with bounces above the mirror while the horizontal dashed lines represent the paths through different GQS.
The 2D figure is drawn in the $y,z$-plane which corresponds to the most probable plane in the distribution of velocities. 

The quantum mirror in the plane $z=0$ produces bounces which constrain the wave function to remain in the half space $z>0$ until it reaches the end of the mirror. This strongly affects the vertical evolution which will be conveniently described by a decomposition on the basis of the eigen-solutions of the Schr\"odinger equation above the quantum reflecting mirror \cite{Nesvizhevsky2015}
\begin{equation}
\chi_n(z)= \frac{\Theta(z)}{\sqrt{\ell_g}} 
\frac{\Ai\left(\frac z{\ell_g}-\lambda_n\right)}
{\Ai^\prime\left(-\lambda_n\right)}  \,.
\label{eigensolutions}
\end{equation}

Properties of the Airy functions $\Ai$ are discussed for example in the NIST Handbook of Mathematical Functions \cite{Olver2010} or in the book by Vallée and Soares \cite{Vallee2004}. $\Theta(z)$ is the Heaviside function expressing perfect quantum reflection; $\ell_g\simeq 5.87 ~\mu$m is the typical length scale associated with the quantum effect in the Earth gravity field (numerical values calculated for $g=g_0$ with $g_0=9.81\,$m/s$^2$ the standard Earth's gravity field)
\begin{equation}
\ell_g = \left(\frac{\hbar^2}{2 m^2 g}\right)^{1/3}  \quad,\quad
\epsilon_g = m g \ell_g  ~.
\label{typicalscales}
\end{equation} 
 $\epsilon_g\simeq 0.602 ~$peV is the energy scale which determines the energy of the $n-$th state $E_n = \lambda_n\epsilon_g $ with $\lambda_n$ the opposite of the $n-$th zero of the Airy function. 
We also introduce typical scales 
for time ($t_g = \frac{\hbar}{\epsilon_g} \approx 1~$ms) 
and velocity ($v_g = g t_g  \approx 1~$cm/s). 
With a radius of 5~cm for the disk, the time spent above it is approximately 50~ms, corresponding to a large number of bounces above the mirror. 
Atoms with a low vertical velocity above the surface are trapped by the combined action of quantum reflection and gravity.

Atoms having a large enough vertical velocity are absorbed by a rough plate placed at some height $z_\mmax $ above the quantum reflecting mirror. Precisely, the absorber selects the states with $n<n_\mmax $ such that the eigen-state $\chi_n$ can pass through the slit \cite{Meyerovich2006,Escobar2014,Dufour2014}. This important point is taken into account in the following by restricting the number of GQS to the range $n\in\left[1,n_\mmax \right]$, which of course limits the number of atoms useful for the measurement. We stress again here that we consider an horizontal polarisation of the photo-detachment laser in order to maximize the number of atoms which pass through the slit formed by the quantum mirror and the absorber. 

After the end of the disk, the atomic wave packet evolves through a free fall down to its annihilation on the detection plate. The free fall acceleration $\gbar$ is deduced from a statistical analysis of annihilated events. The key reason for the improvement of accuracy of the quantum experiment with respect to the classical one is the fact that the quantum interference pattern on the detector contains much more information on the value of $\gbar$ than the classical pattern.

We will neglect the effect of initial position dispersion $\Delta x=\Delta y\simeq0.5~\mu$m with respect to the macroscopic dimensions of the whole setup. The situation is completely different for the dispersion $\Delta z$ which has the same value but will be taken into account carefully in the following, as $\Delta z$ is not small in comparison for the parameters $\ell_g$ and $h$ of interest for vertical evolution. We will also use the fact that horizontal velocities are preserved during the whole experiment to relate them to the measured positions $(X,Y,Z,T)$ in space and time of the annihilation event. 

To write the associated relations, we introduce notations for projections of vectors in the horizontal plane 
\begin{equation}
\obv=\left(v_x , v_y \right)= \frac {\obR}T   \quad , \quad 
\obR=\left(X , Y \right)  ~.
\label{horizontalprojections}
\end{equation} 
We also deduce the positions of the atom at the end of the quantum mirror, just before the free fall 
\begin{equation}
t=\frac d{\onv}=\frac {Td}{\onR} 
\quad , \quad \onv=\vert\obv\vert 
\quad , \quad \onR=\vert\obR\vert~.
\label{horizontalpositions}
\end{equation} 

\section{Interferences above the quantum mirror}
\label{sec:interference}

We now detail the evolution of the density matrix above the quantum mirror. We suppose that most atoms survive their flight above the reflecting surface, thus neglecting the annihilation on the disk due to nearly perfect quantum reflection for atoms having a small orthogonal velocity above the quantum  reflecting mirror. 

\subsection{Evolution above the quantum mirror}

The pure state $\ket{\Psi_{\bq,0}}$ corresponding to a well-defined recoil in Eq.\eqref{initiamdensitymatrix} and taken at time 0, before evolution above the quantum mirror, can be written with factorized horizontal and vertical evolutions (the former written in terms of horizontal projections \eqref{horizontalprojections}, the latter in terms of $\delta z_0=z_0-h$)
\begin{equation}
\begin{aligned}
&\Psi_{\bq,0}(\br_0)=\phi_0(\obr_0)~\psi_0(z_0)\,, \\
& \phi_0(\obr_0)=\frac{1}{\sqrt{2\pi} \zeta}
   \exp\left(-\frac{\obr_0^2}{4\zeta^2}
+\frac{\imath\,\obq . \obr_0}{\hbar}\right) \,, \\
&\psi_0(z_0)=\sqrt{\frac{1}{\sqrt{2\pi} \zeta}}
    \exp\left(-\frac{\delta z_0^2}{4\zeta^2}
+\frac{\imath\,q_z \delta z_0}{\hbar}\right) \,.
\end{aligned}
\label{wavepacketevolution0}
\end{equation} 

In the configuration described in \S\ref{sec:reminders}, the evolution over the mirror is conveniently represented in the momentum representation for the horizontal variables as horizontal momenta are conserved quantities 
\begin{equation}
\widetilde{\phi}_0(\obp)=
\frac{1}{\sqrt{2\pi} \Delta p}
\exp\left(-\frac{(\obp-\obq)^2}{4\Delta p^2} \right) \,.
\end{equation}
We will write below the expression of the wave packet after an evolution for a time $t$ spent above the quantum reflecting mirror with $t$ related to horizontal velocities by Eq.\eqref{horizontalpositions}.
In other words, the horizontal evolution is trivial when written in the momentum representation. 

In contrast, the vertical evolution is strongly affected by quantum reflection and is conveniently represented by a decomposition over the GQS (with energy $\lambda_n\epsilon_g$)
\begin{equation}
\psi_{\bq,t}(z)=\sum\limits_n c_n(q_z)  \chi_{n}(z) 
\exp\left(-\frac {\imath\lambda_n\epsilon_g t}\hbar\right) \,.
\label{wavepacketevolution}
\end{equation}
The amplitudes $c_n$ depend on the vertical recoil $q_z$ and can be calculated as overlap integrals (the lower bound in the integral is set at 0 because the functions $\chi_n(z)$ contain an Heaviside function $\Theta(z)$)
\begin{equation}
c_n(q_z) = \int_{0}^{+\infty} \psi_0(z_0) \chi_n(z_0)^* \md z_0 \,.
\label{overlapintegral}
\end{equation}
The integral \eqref{overlapintegral} has a good analytical approximation when the dispersion $\zeta$ is small with respect to $h$ so that the lower bound in the integral \eqref{overlapintegral} can be changed to $-\infty$.
In the following we use the exact integral \eqref{overlapintegral}. 

Among the $N=1000$ initial $\Hbar$ atoms, some of them are lost in the absorber which reduces the number of detected atoms $N_c$ to a value which can be deduced from the quadratic sum of the coefficient $c_n$
\begin{equation}
N_c = N ~ \int d\Omega~\varpi(\hbq)\sum\limits_n 
\left\vert c_n(q_z)\right\vert^2 ~. 
\label{eq:N-photo-detachment}
\end{equation} 
With the numbers used here ($f=20\,$kHz, $\delta E=10~\mu$eV and $n_\mmax =1000$), we deduce that 26\% of the initial atoms pass through the slit and are detected ($N_c=260$ when $N=1000$). 

We can now write a density matrix having the same form as in \eqref{initiamdensitymatrix} evaluated at time $t$
at the end of the disk
\begin{equation}
\rho_t = \int \,\varpi(\hbq) \,\md\Omega 
\ket{\Psi_{\bq,t}}\bra{\Psi_{\bq,t}} ~, 
\label{enddiskdensitymatrix}
\end{equation} 
This density matrix contains all the information which will be needed in the following to evaluate precisely the extent to which the incoherent part may affect the interference fringes with respect to the case without photo-detachment. It will be used as the basis of the study of free fall in the next section \S\ref{sec:patterndetector}. 

\subsection{Momentum distribution at the end of the disk}

Before embarking in this complete study, we want to confirm that the interferences are still present when taking the recoil into account. 

To this aim, we calculate the momentum distribution at the end of the mirror which was shown in \cite{Crepin2019} to contain interferences
\begin{equation}
\begin{aligned}
&\Pi_t(\bp) = \int \varpi(\hbq)\md\Omega \,
\frac{\exp\left(-\frac{(\obp-\obq)^2}{2\Delta p^2}\right) }{2\pi \Delta p^2}
\left\vert \widetilde{\psi}_{\bq,t}(p_z)\right\vert^2 \,,  \\
&\widetilde{\psi}_{\bq,t}(p_z) = \sum\limits_{n}  
c_n(q_z)  \widetilde{\chi}_{n}(p_z)
\exp\frac {-\imath\lambda_n t}{t_g} \,,  
\end{aligned}
\label{eq:momentum-distribution}
\end{equation} 
where the symbols $\widetilde{\psi}_{\bq,t}$ and $\widetilde{\chi}_n$ represent the Fourier transforms of the functions ${\psi}_{\bq,t}$ and  ${\chi}_n$ respectively.

\begin{figure}[t!]
\centering
\includegraphics[scale=0.55]{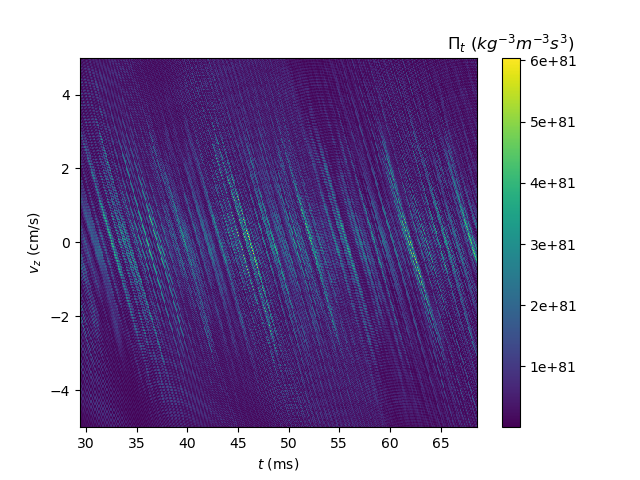}
\caption{Momentum distribution $\Pi_t$ represented as a function of velocity $v_z=p_z/m$ and time $t$ spent above the mirror, for the parameters $f=20\,$kHz, $\delta E=10~\mu$eV and $n_\mmax =1000$.}
\label{fig:Pi_d-vz-photo-detachment}
\end{figure} 

The distribution \eqref{eq:momentum-distribution} is represented in Fig.\ref{fig:Pi_d-vz-photo-detachment} as a function of velocity $v_z$ and time $t$  spent above the mirror. The pattern clearly reveals interference fringes due to interference between the paths corresponding to different GQS above the quantum mirror. 
Using a far-field approximation for describing the free fall after the end of the mirror, one would deduce the interference pattern on the detector directly from this velocity distribution as was done in \cite{Crepin2019}. However, this approximation is no longer valid for the parameters of the present paper and we use below a more rigorous way based on the full information contained in the density matrix \eqref{enddiskdensitymatrix}. 

\section{Interference pattern on the detection plates}
\label{sec:patterndetector}

The interference pattern produced at the end of the quantum mirror is translated by the free fall period into another one which is read out from the positions in space and time of annihilation events of $\Hbar$ at the detection plates. For simplicity, we assume that $\Hbar$ are annihilated with 100\% probability at the detector, thus disregarding quantum reflection there as the kinetic energy of atoms is large after a macroscopic free fall height.

\subsection{Derivation of the annihilation current}

The free fall time $\tau$ is the difference between the full evolution time $T$ and the time $t$ spent above the disk
\begin{equation}
\tau = T - t  ~. 
\end{equation} 
For a pure wave function $\ket{\Psi_{\bq,t}}$ at time $t$, the pure wave function $\ket{\Psi_{\bq,T}}$ at time $T$ after the free fall time $\tau$ is conveniently described by a quantum propagator.
The propagator is represented as follows in the position representation with position $z$ at time $t$ and $Z$ at time $T$ 
\begin{equation}
\begin{aligned}
&\Psi_{\bq,T}(Z) = \int \md z {K}_\tau^g(z,Z) \Psi_{\bq,t}(z) ~, \\
&K_\tau^g(z,Z)=  \sqrt{\frac{m}{2\imath\pi\hbar \tau} }
\exp\left(\frac{\imath m\,s_\tau^g(z,Z)}{\hbar}\right) ~, \\
&s_\tau^g(z,Z)= 
\frac{\left( Z-z\right) ^{2}}{2\tau}
-\frac{g\tau\left( Z+z\right) }{2}-\frac{g^2\tau^3}{24} ~.  
\end{aligned}
\label{propagator}
\end{equation} 
This expression has a nice interpretation as $ms_\tau^g(z,Z)$ is simply the action on the classical trajectory from $z$ to $Z$ on a time $\tau$ \cite{Storey1994}. 

The quantum propagator $K_\tau^g$ in presence of the gravity field can easily be rewritten in terms of the same quantity $K_\tau^0$ evaluated in the absence of gravity and of a change of final altitude accounting for the mean free fall height  
\begin{equation}
\begin{aligned}
&K_\tau^g(z,Z) = \exp\left(-\imath\Phi_\tau(Z)\right)
K_\tau^0\left(z,Z^\prime\right)  ~, \\
&\Phi_\tau(Z)=\frac{mg\tau}{\hbar}\left(Z+\frac{g\tau^2}{6}\right) 
\quad,\quad Z^\prime=Z+\frac{g\tau^2}{2} ~.
\end{aligned}
\end{equation} 
This formula is interesting as it dissociates the description of gravity by the mean free fall height $\tfrac{1}{2}g\tau^2$ and that of diffraction by the propagator $K_\tau^0$ containing only the first term 
$s_\tau^0\left(z, Z^\prime\right) = \frac{\left( Z^\prime-z\right) ^{2}}{2\tau}$ 
in the full action in the third line of \eqref{propagator}. The treatment of gravity is thus compatible with the equivalence principle while that of diffraction depends on the value of $m/\hbar$. 
 
The global phase factor $\Phi_\tau(Z)$ does not depend on $z$ 
and thus goes out of the propagation integral in the first line of \eqref{propagator}. It would be irrelevant for calculating the probability density at the detector, but we calculate here the annihilation probability current $J$ at the detector. $J$ is a number of particle per unit of time and per unit of surface and a function of the space-time positions $X,Y,Z,T$ of the annihilation event with $Z=-H$ fixed on the detection plane. In the configuration sketched in Fig.\ref{fig:quantum-design} where the atoms have a negative vertical velocity at the detector, $J$ has to be defined as the opposite of the $z-$component of the current vector, with the expression
\begin{equation}
\begin{aligned}
&J =\frac{m^2}{\tau^2}\int \varpi(\hbq)\md\Omega \,
\frac{\exp\left(-\frac{(\obp-\obq)^2}{2\Delta p^2}\right) }%
{2\pi \Delta p^2}
j_{\bq}~,\\
&j_{\bq}=- \Re
\left(\psi _\bq^\ast(Z) \frac{\hbar}{\imath m}\partial_Z \psi _\bq (Z) \right)  ~. 
\label{eq:current-photo-detachment} 
\end{aligned}
\end{equation} 
The parameter $T$ has been omitted for readability and 
the relations written above between variables at the end of the disk and positions $X,Y,Z,T$ of the annihilation event have also been left implicit. 

An explicit expression of the current is then obtained by
rewriting the wave function $\ket{\Psi_{\bq,T}}$
\begin{equation}
\begin{aligned}
&\Psi_{\bq,T}(Z) = \exp\left(-\imath\Phi_\tau(Z)\right)  \sqrt{\frac{m}{2\imath\pi\hbar \tau} } \times  \\
&\qquad \int \md z
\exp\left(\frac{\imath m\left( Z^\prime-z\right) ^2}{2\hbar\tau}\right) \Psi_{\bq,t}(z)
 ~, 
\end{aligned}
\label{eq:psi-diffraction-photo-detachment} 
\end{equation} 
as well as its gradient versus $Z$
\begin{equation}
\begin{aligned}
&\frac{\hbar}{\imath m}\partial_Z\Psi_{\bq,T}(Z) = \exp\left(-\imath\Phi_\tau(Z)\right)  \sqrt{\frac{m}{2\imath\pi\hbar \tau} } \times  \\
&\quad\int \md z V(z,Z)\exp\left(\frac{\imath m\left( Z^\prime-z\right) ^2}{2\hbar\tau}\right) \Psi_{\bq,t}(z) ~, \\
&V(z,Z) = \frac{Z-z}{\tau}-\frac{g\tau}{2} ~.
\end{aligned}
\label{eq:gradpsi-diffraction-photo-detachment} 
\end{equation} 
Here, $V$ is the (negative) vertical velocity at the detector calculated for the classical motion from the altitude $z$ at time $t$ to $Z$ at time $T$. Note that it depends on $z$ and cannot be taken out of the integral in \eqref{eq:gradpsi-diffraction-photo-detachment}.

Equation \eqref{eq:psi-diffraction-photo-detachment} describes quantum diffraction with the effect of gravity accounted for by the altitude change $Z\to Z^\prime$. It allows to discuss easily the approximation of far-field diffraction, as the integrals in \eqref{eq:psi-diffraction-photo-detachment} and \eqref{eq:gradpsi-diffraction-photo-detachment} are restricted to the interval $z\in\left[0,z_\mathrm{max}\right]$. The far-field limit can be used and the current expressed in terms of the momentum distribution at the end of the disk when $\tau$ is longer than a time of the order of $\frac{m z_\mmax ^2}{h}$ (that is also when the distance $\overline{\overline{v}}\tau$ is longer than $\frac{z_\mmax ^2}{\lambda_\mathrm{dB}}$ with $\lambda_\mathrm{dB}=\frac{h}{m\overline{\overline{v}}}$ the associated de Broglie wavelength). With the numbers used in \cite{Crepin2019}, this approximation gave a fairly good result. With the numbers in the present paper in contrast, the approximation can no longer be used, and we have to proceed with the more demanding numerical evaluation of the formula \eqref{eq:current-photo-detachment}, with the wave function and its gradient given by \eqref{eq:psi-diffraction-photo-detachment} and \eqref{eq:gradpsi-diffraction-photo-detachment} respectively.

\subsection{Discussion of the annihilation current}

\begin{figure}[t!]
\centering
\includegraphics[scale=0.55]{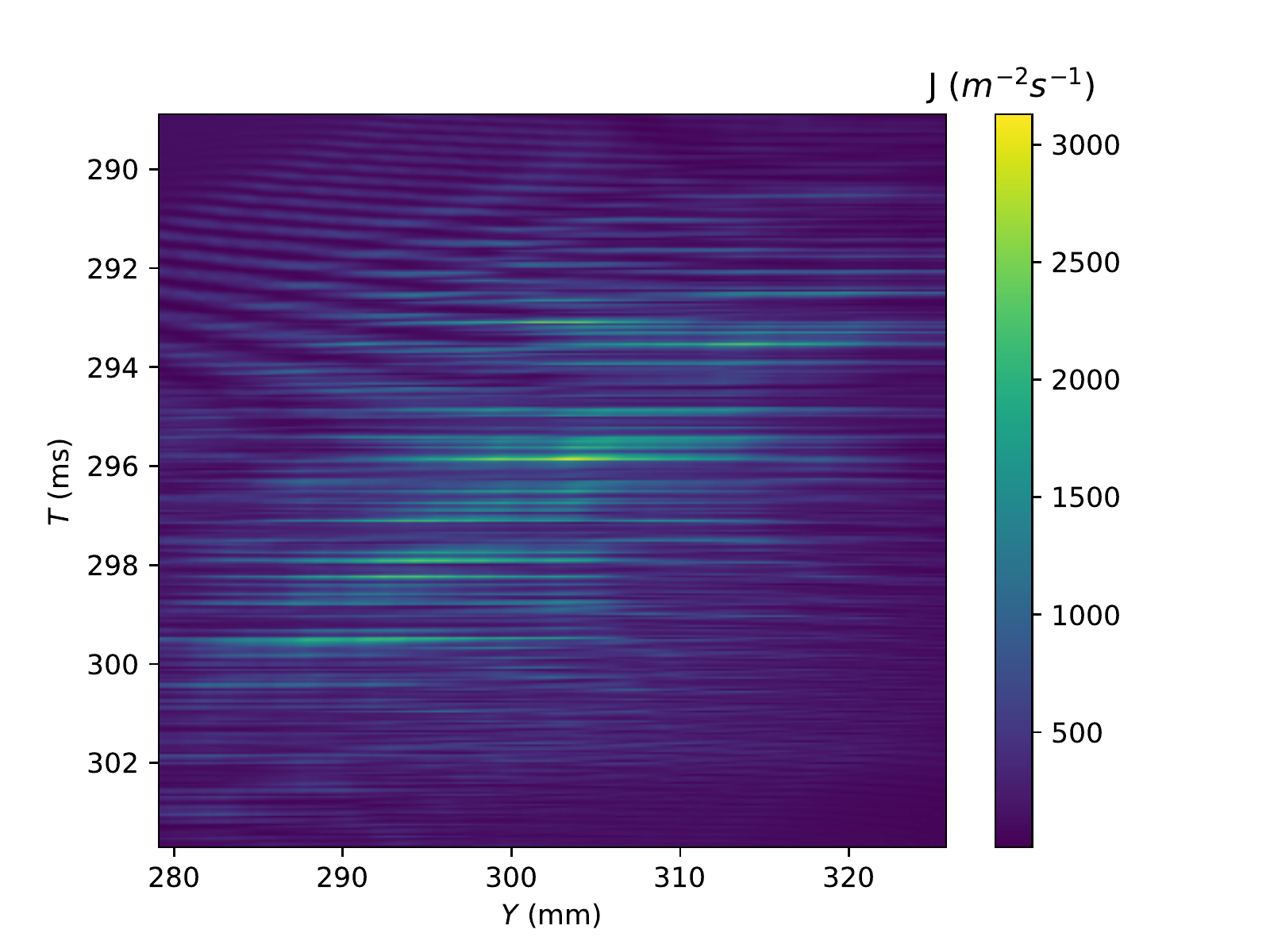}
\caption{Probability current density $J(Y,T)$ on the detection plate, for $Z=-H$ and $X=0$. Parameters: $f=20~$kHz, $\delta E=10~\mu$eV, $n_\mmax =1000$.}
\label{fig:current-quantum-photo-detachment-2D}
\end{figure} 

We represent in the Fig.\ref{fig:current-quantum-photo-detachment-2D} the current \eqref{eq:current-photo-detachment} versus $Y$ and $T$ for $X=0$ and $Z=-H$. The center of the figure is at the classical position corresponding to a null vertical velocity at the end of the disk $(Y,T)=(d+v_\mr \tau,d/v_\mr +\tau)\simeq\,$(302 mm, 296 ms). The figure clearly shows interference fringes along mainly horizontal lines, which means that the interference is now essentially encoded on the time of arrival. The pattern is organized around a most probable line corresponding to the diagonal $Y=v_\mr T$.  

\begin{figure*}[t!]
\centering
\includegraphics[width=.9\linewidth]{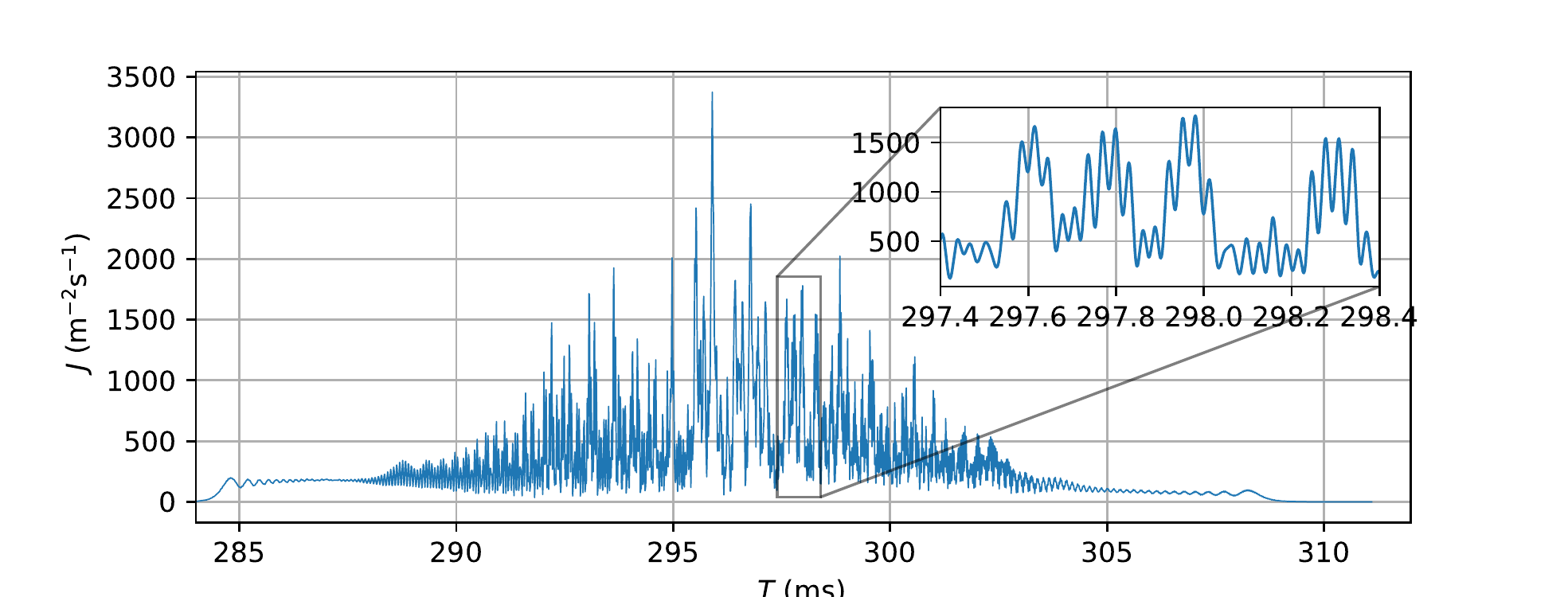}
\caption{Probability current density $J$ in $m^{-2}s^{-1}$ on the detection plate, by fixing $Z=-H$, $X=0$ and $Y=v_\mr T$.}
\label{fig:current-quantum-photo-detachment}
\end{figure*}

Details of the interference pattern are emphasized by cutting the 2D distribution in Fig.\ref{fig:current-quantum-photo-detachment-2D} along the most probable line $Y=v_\mr T$. For the reasons just explained, we plot it as a function of time $T$ around the classical center 296 ms in Fig.\ref{fig:current-quantum-photo-detachment}.
A zoom of Fig.\ref{fig:current-quantum-photo-detachment} is represented in the inset with the aim of showing that the pattern is indeed an interference signal, with however a complex structure due to the large number of frequencies involved in its expression. 

\section{Estimation of the uncertainty}
\label{sec:uncertainty}

We now have all the necessary ingredients to estimate the precision of the experiment and compare it with the purely classical design.
We use two statistical methods to estimate the parameter $g$ and deduce a variance for this estimation, the Monte-Carlo numerical method and the Cramer-Rao analytical method. 
Atoms annihilate on an horizontal detector plate at $Z=-H$ with the event characterized by space positions $X_i,Y_i$ and time $T_i$.

\subsection{Monte-Carlo simulation}

In the Monte-Carlo simulation, detection events are generated directly from the expression \eqref{eq:current-photo-detachment} of the current $J$ at detection. We choose randomly $N_c$ detection events in the probability distribution $J$ corresponding to the value $g_0=9.81\,$m/s$^2$. We consider that the random draw $\mathcal{D}$ of $N_c$ detection events $\left(X_i,Y_i,T_i\right)$ simulates the output of one experiment.

The atomic evolution from the source to the detector doesn't depend on the azimuth angle $\Phi$ though the initial momentum distribution depends on $\Phi$. We may take benefit of this symmetry by using cylindrical coordinates, each event corresponding to a horizontal radius $\onR_i$ and an azimuth angle $\Phi_i$.
We can even produce an equivalent 2D analysis gathering all information by summing over angles $\Phi$ folded on the value $\frac\pi2$. 

\begin{figure}[b!]
\centering
\includegraphics[scale=0.55]{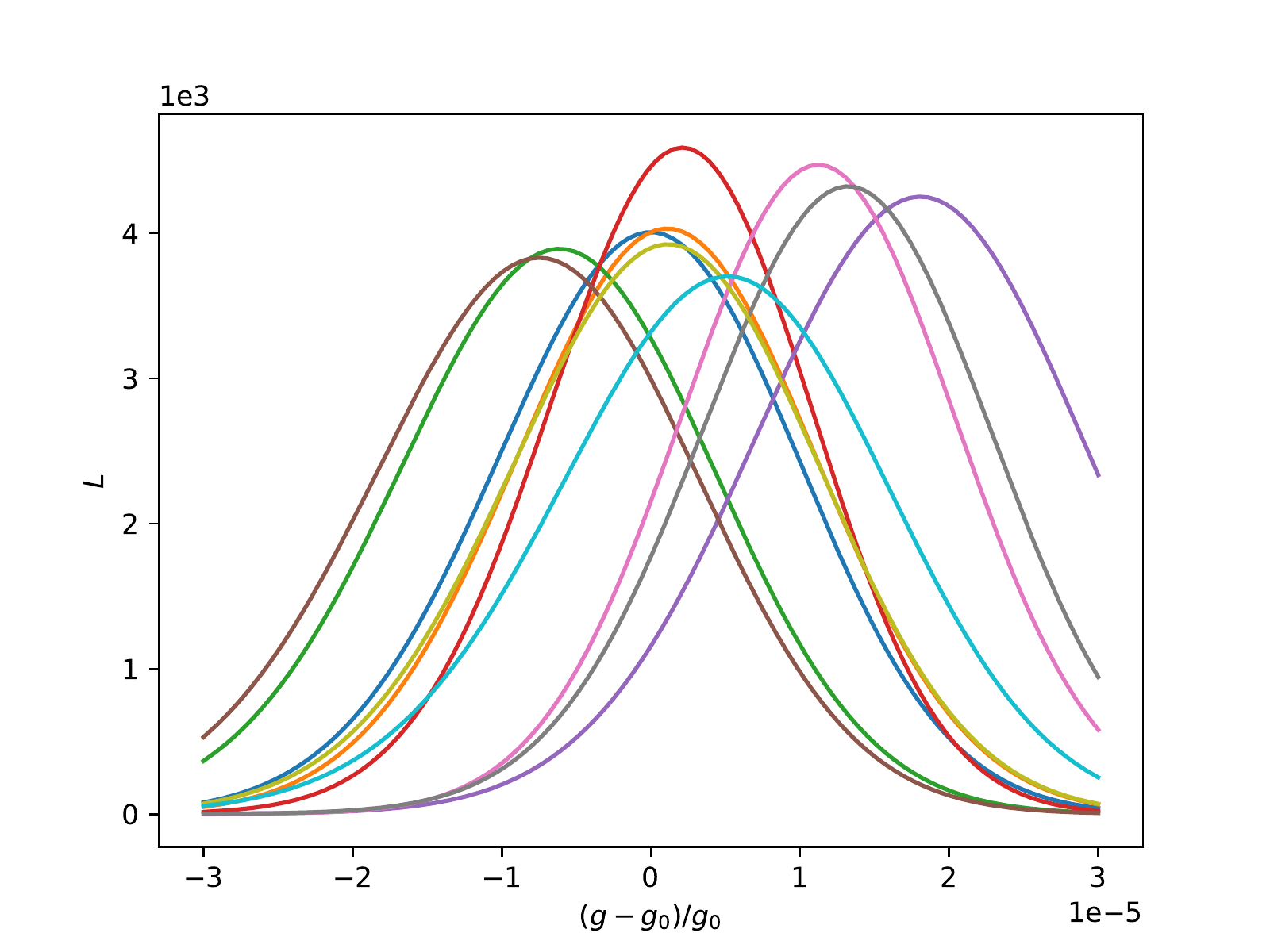}
\caption{Likelihood functions $L$ calculated for 10 random draws of $N_c=260$ atoms.}
\label{fig:likelihoods-quantum-photo-detachment}
\end{figure}

\begin{figure}[b!]
\centering
\includegraphics[scale=0.55]{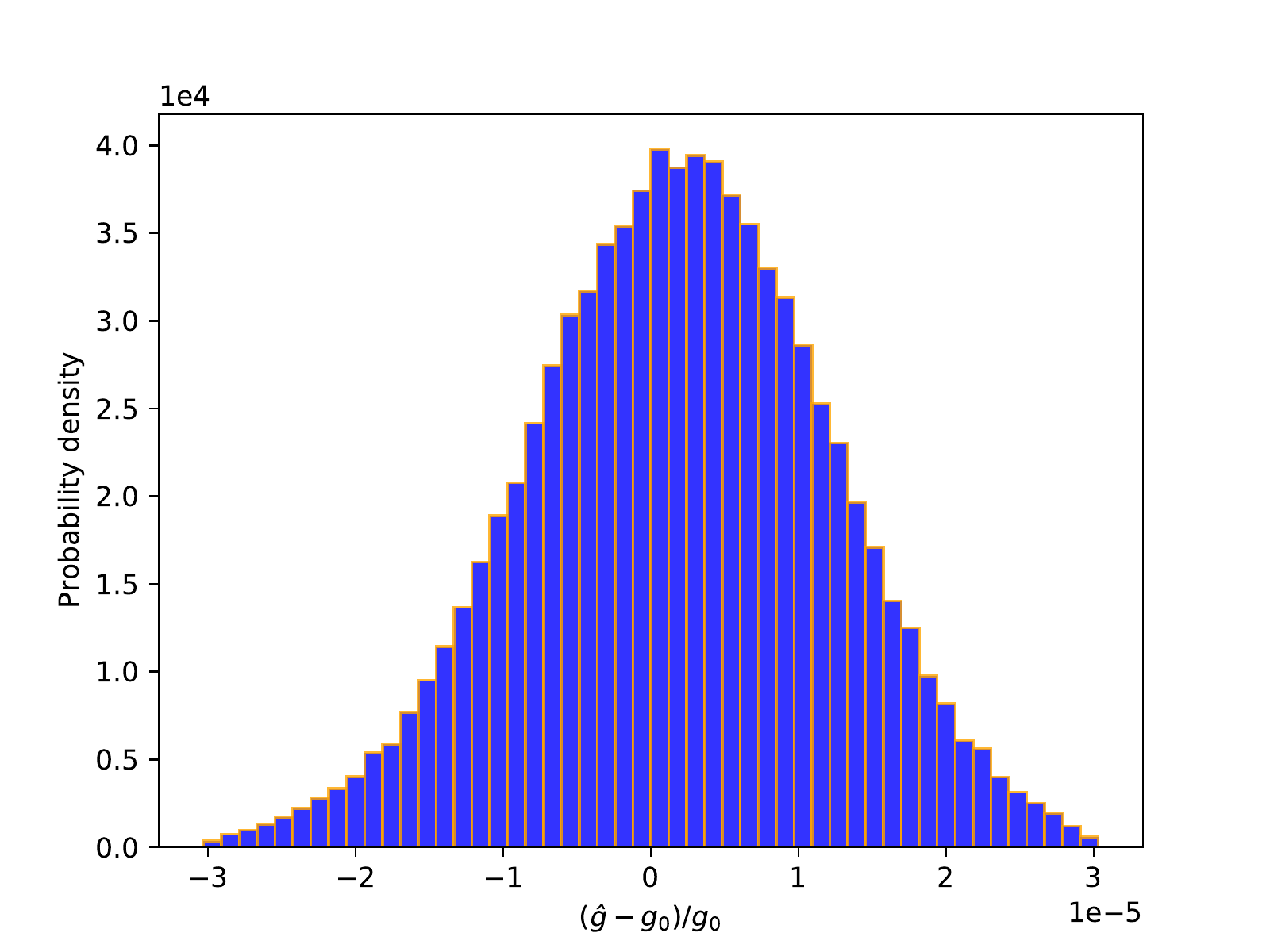}
\caption{Normalized histogram of the relative variation $(\widehat{g}-g_0)/g_0$ of the estimator $\widehat{g}$ obtained by repeating 100000 times the Monte-Carlo simulation on $N_c=260$ atoms.}
\label{fig:histogram-quantum-photo-detachment}
\end{figure}

The Fig.\ref{fig:likelihoods-quantum-photo-detachment} shows samples of likelihoods defined in \cite{Rousselle2022NJP}, corresponding to 10 independent random draws. The colors have no meaning, they only allow one to distinguish the different likelihood functions. The horizontal axis scales as $\frac{g-g_0}{g_0}$, in the interval  $\left[-3\cdot10^{-5};3\cdot10^{-5}\right]$.
The likelihoods have Gaussian shapes with nearly equal variances, the main difference from one to the other being the position of the maximum. 
We use here the max likelihood method to get an estimator $\widehat{g}$ of the parameter $g$ as would be done in the data analysis of the experiment. 
In order to give a robust estimation of the variance, we repeat the full procedure for $M$ different random draws of the $N_c$ points. The histogram shown in Fig.\ref{fig:histogram-quantum-photo-detachment} corresponds to $M=100000$ such draws of $N_c$ points. 

This histogram has a Gaussian shape, with a dispersion corresponding approximately to the average dispersion of the likelihood functions. From this histogram, we extract the average $\mu_g$ and the dispersion $\sigma_g^\MC $ of the estimator $\widehat{g}$, from which we calculate the relative uncertainty 
\begin{equation}
  \frac{\sigma_g^\MC}{g_0} \simeq 1.0 \cdot 10^{-5}~.
  \label{eq:monte-carlo-uncertainty}
\end{equation}

In the case without photo-detachment, with the same design parameters and an initial horizontal kick $v_0=1.02$ m/s, about 995 atoms would be detected since the dispersion of initial vertical velocity is much smaller and a relative uncertainty $\sigma_g^\MC /g_0 \approx 5.8 \cdot 10^{-6}$ would be obtained.
Taking into account the fact that the number of detected atoms is reduced by a factor 0.26 by the spread of photo-detachment recoil, we see that the latter only slightly decreases the precision per detected atom. In other words, the photo-detachment degrades the precision not so much as a result  of blurring of the interference figure, but rather because about 74\% of the initial atoms are absorbed in the slit above the quantum mirror.

\subsection{Cramer-Rao method and statistical efficiency} 

We now compare this result with the Cramer-Rao bound $\sigma_g^\CR $  deduced from the Fisher information $\mathcal{I}_g$ for $g$-dependence of the event distribution \cite{Frechet1943,Cramer1999,Refregier2004}
\begin{equation}
\begin{aligned}
&\sigma_g^\CR =\frac{1}{\sqrt{N~\mathcal{I}_g}}\,,\\
\mathcal{I}_g &= \int \md X \md Y \md T ~
\frac{\left(\partial_g J_g(X,Y,T)\right)^2}{J_g(X,Y,T)}\,, 
\end{aligned}
\label{eq:cramer-rao-bound}
\end{equation}
where $J_g$ is the current \eqref{eq:current-photo-detachment} calculated for a running value $g$ of the free fall acceleration. 
We note that $N$ is the initial number of atoms and that $J_g$ accounts for the absorption of atoms in the slit above the mirror. In other words, $\mathcal{I}_g$ is the Fisher information per incident atom, with \eqref{eq:cramer-rao-bound} accounting for the loss of around 74\% of the atoms which do not bring any information on the value of $g$. 

The Cramer-Rao dispersion \eqref{eq:cramer-rao-bound} corresponds to an optimal estimation of the parameter \cite{Cramer1999,Refregier2004}.
In the present problem, the relative uncertainty obtained from \eqref{eq:cramer-rao-bound} is slightly smaller than the one \eqref{eq:monte-carlo-uncertainty} calculated by the Monte Carlo simulation
\begin{equation}
 \frac{\sigma_g^\CR}{g_0} \simeq0.98 \cdot 10^{-5} \,.
\label{eq:cramer-rao-uncertainty}
\end{equation}
This means that the statistical efficiency \cite{Cramer1999}, defined as the ratio between the variances $(\sigma_g^\MC)^2$ and  $(\sigma_g^\CR)^2$, is good ($e\approx 0.96$ close to 1).
From an experimental point of view, a good efficiency means that the unique random draw to be obtained from one experiment is expected to be representative of the variety of results obtained from different random draws in the numerical simulations.

\section{Conclusion}
\label{sec:conclusion}

In this article, we have studied in details the quantum design of the GBAR experiment, by taking into account the photo-detachment recoil. The idea was suggested in \cite{Crepin2019}, without accounting for the photo-detachment.
Furthermore, a far-field approximation was used there for calculating the free fall propagation, while a more accurate quantum propagator method was used here. 

We have described precisely the quantum evolution of the matter wave in this new experimental setup by first studying the $\Hbar$ atoms bouncing on a material mirror, due to high quantum reflection upon the Casimir-Polder potential appearing in the vicinity of the surface. This led to the building of quantum interferences between the paths corresponding to the different Gravitational Quantum States above the mirror. 

This interference pattern at the end of the mirror was then revealed by a free fall down to a detection plate placed at a fixed macroscopic altitude below the quantum mirror. Information on the value of $g$ is encoded there in the distribution of time and space positions of the annihilation events.
The calculations were done here by using the information contained in the full density matrix, without the simplifications based on the far-field diffraction approximation. 

The calculations were done for a superposition of thousand states, which corresponds to about 26\% of atoms detected with the parameters used for the simulations. A  good agreement was reached between the two statistical analysis performed with Monte Carlo simulation and Cramer-Rao estimation. This means that the statistical efficiency is good, a nice result regarding the complexity of the interference pattern and the relatively low number of atoms probing this pattern. 

The main conclusion of this paper is that the photo-detachment process degrades the precision mainly because the spreading of the vertical velocity leads to a loss of atoms absorbed in the slit above the quantum mirror. Meanwhile, the blurring of the interference pattern which could have affected the precision of the quantum experiment is not significant. 
In the end, the relative uncertainty is spectacularly improved when going from the classical timing measurement to the quantum one. With the set of parameters considered here, we get a relative accuracy of $3.3\cdot10^{-2}$ for the classical design and about $10^{-5}$ for the quantum one, thus improving the uncertainty by more than 3 orders of magnitude. The main reason for this improvement is that the quantum interference pattern contains much more information on the value of $g$ than the Gaussian-like classical pattern. Fine details act as thin graduations that make it easier to observe small changes of the probability current distribution when the estimated parameter varies.

Further advances would be necessary for a fully realistic estimation of the uncertainty. In particular, the quantum reflection on the mirror and on the detection plate should be accounted for \cite{Dufour2013} and the effect of  position resolution at the detection plate \cite{Radics2019} should be included in the analysis as it can blur the finest fringes. It would also be necessary to treat the effect of the shifts of Gravitational Quantum States on the Casimir-Polder potential, an effect which has already been calculated with the required accuracy \cite{Crepin2017,Crepin2019d}. These advances will be needed for a precise analysis of the quantum experiment when it will be performed but they will not change the main result obtained in this paper, namely that the quantum design leads to a better accuracy than the classical one.

\paragraph*{Acknowledgements}
We thank our colleagues in the GBAR collaboration \url{https://gbar.web.cern.ch/} for insightful discussions, in particular P.P. Blumer, C. Christen, P.-P. Cr\'epin, P. Crivelli, P. Debu, A. Douillet, N. Garroum, L. Hilico, P. Indelicato, G. Janka, J.-P. Karr, L. Liszkay, B. Mansouli\'e, V.V. Nesvizhevsky, F. Nez, N. Paul, P. P\'erez, C. Regenfus, F. Schmidt-Kaler, A.Yu. Voronin, S. Wolf.
This work was supported by the Programme National GRAM of CNRS/INSU with INP and IN2P3 co-funded by CNES and by Agence Nationale pour la Recherche, Photoplus project No. ANR-21-CE30-0047-01.  

\bibliography{gbarqu}

\begin{thebibliography}{56}%
\makeatletter
\providecommand \@ifxundefined [1]{%
 \@ifx{#1\undefined}
}%
\providecommand \@ifnum [1]{%
 \ifnum #1\expandafter \@firstoftwo
 \else \expandafter \@secondoftwo
 \fi
}%
\providecommand \@ifx [1]{%
 \ifx #1\expandafter \@firstoftwo
 \else \expandafter \@secondoftwo
 \fi
}%
\providecommand \natexlab [1]{#1}%
\providecommand \enquote  [1]{``#1''}%
\providecommand \bibnamefont  [1]{#1}%
\providecommand \bibfnamefont [1]{#1}%
\providecommand \citenamefont [1]{#1}%
\providecommand \href@noop [0]{\@secondoftwo}%
\providecommand \href [0]{\begingroup \@sanitize@url \@href}%
\providecommand \@href[1]{\@@startlink{#1}\@@href}%
\providecommand \@@href[1]{\endgroup#1\@@endlink}%
\providecommand \@sanitize@url [0]{\catcode `\\12\catcode `\$12\catcode
  `\&12\catcode `\#12\catcode `\^12\catcode `\_12\catcode `\%12\relax}%
\providecommand \@@startlink[1]{}%
\providecommand \@@endlink[0]{}%
\providecommand \url  [0]{\begingroup\@sanitize@url \@url }%
\providecommand \@url [1]{\endgroup\@href {#1}{\urlprefix }}%
\providecommand \urlprefix  [0]{URL }%
\providecommand \Eprint [0]{\href }%
\providecommand \doibase [0]{https://doi.org/}%
\providecommand \selectlanguage [0]{\@gobble}%
\providecommand \bibinfo  [0]{\@secondoftwo}%
\providecommand \bibfield  [0]{\@secondoftwo}%
\providecommand \translation [1]{[#1]}%
\providecommand \BibitemOpen [0]{}%
\providecommand \bibitemStop [0]{}%
\providecommand \bibitemNoStop [0]{.\EOS\space}%
\providecommand \EOS [0]{\spacefactor3000\relax}%
\providecommand \BibitemShut  [1]{\csname bibitem#1\endcsname}%
\let\auto@bib@innerbib\@empty
\bibitem [{\citenamefont {Hori}\ and\ \citenamefont {Walz}(2013)}]{Hori2013}%
  \BibitemOpen
  \bibfield  {author} {\bibinfo {author} {\bibfnamefont {M.}~\bibnamefont
  {Hori}}\ and\ \bibinfo {author} {\bibfnamefont {J.}~\bibnamefont {Walz}},\
  }\bibfield  {title} {\bibinfo {title} {Physics at {{CERN}}'s antiproton
  decelerator},\ }\href
  {https://doi.org/https://doi.org/10.1016/j.ppnp.2013.02.004} {\bibfield
  {journal} {\bibinfo  {journal} {Progress in Particle and Nuclear Physics}\
  }\textbf {\bibinfo {volume} {72}},\ \bibinfo {pages} {206} (\bibinfo {year}
  {2013})}\BibitemShut {NoStop}%
\bibitem [{\citenamefont {Bertsche}\ \emph {et~al.}(2015)\citenamefont
  {Bertsche}, \citenamefont {Butler}, \citenamefont {Charlton},\ and\
  \citenamefont {Madsen}}]{Bertsche2015}%
  \BibitemOpen
  \bibfield  {author} {\bibinfo {author} {\bibfnamefont {W.}~\bibnamefont
  {Bertsche}}, \bibinfo {author} {\bibfnamefont {E.}~\bibnamefont {Butler}},
  \bibinfo {author} {\bibfnamefont {M.}~\bibnamefont {Charlton}},\ and\
  \bibinfo {author} {\bibfnamefont {N.}~\bibnamefont {Madsen}},\ }\bibfield
  {title} {\bibinfo {title} {Physics with antihydrogen},\ }\href
  {https://doi.org/10.1088/0953-4075/48/23/232001} {\bibfield  {journal}
  {\bibinfo  {journal} {Journal of Physics B: Atomic, Molecular and Optical
  Physics}\ }\textbf {\bibinfo {volume} {48}},\ \bibinfo {pages} {232001}
  (\bibinfo {year} {2015})}\BibitemShut {NoStop}%
\bibitem [{\citenamefont {Charlton}\ \emph {et~al.}(2017)\citenamefont
  {Charlton}, \citenamefont {Mills},\ and\ \citenamefont
  {Yamazaki}}]{Charlton2017}%
  \BibitemOpen
  \bibfield  {author} {\bibinfo {author} {\bibfnamefont {M.}~\bibnamefont
  {Charlton}}, \bibinfo {author} {\bibfnamefont {A.}~\bibnamefont {Mills}},\
  and\ \bibinfo {author} {\bibfnamefont {Y.}~\bibnamefont {Yamazaki}},\
  }\bibfield  {title} {\bibinfo {title} {Special issue on antihydrogen and
  positronium},\ }\href {https://doi.org/10.1088/1361-6455/aa75d8} {\bibfield
  {journal} {\bibinfo  {journal} {Journal of Physics B: Atomic, Molecular and
  Optical Physics}\ }\textbf {\bibinfo {volume} {50}},\ \bibinfo {pages}
  {140201} (\bibinfo {year} {2017})}\BibitemShut {NoStop}%
\bibitem [{\citenamefont {Yamazaki}(2020)}]{Yamazaki2020}%
  \BibitemOpen
  \bibfield  {author} {\bibinfo {author} {\bibfnamefont {Y.}~\bibnamefont
  {Yamazaki}},\ }\bibfield  {title} {\bibinfo {title} {Cold and stable
  antimatter for fundamental physics},\ }\href
  {https://doi.org/10.2183/pjab.96.034} {\bibfield  {journal} {\bibinfo
  {journal} {Proceedings of the Japan Academy, Series B}\ }\textbf {\bibinfo
  {volume} {96}},\ \bibinfo {pages} {471} (\bibinfo {year} {2020})}\BibitemShut
  {NoStop}%
\bibitem [{\citenamefont {Bondi}(1957)}]{Bondi1957}%
  \BibitemOpen
  \bibfield  {author} {\bibinfo {author} {\bibfnamefont {H.}~\bibnamefont
  {Bondi}},\ }\bibfield  {title} {\bibinfo {title} {Negative mass in general
  relativity},\ }\href {https://doi.org/10.1103/RevModPhys.29.423} {\bibfield
  {journal} {\bibinfo  {journal} {Reviews of Modern Physics}\ }\textbf
  {\bibinfo {volume} {29}},\ \bibinfo {pages} {423} (\bibinfo {year}
  {1957})}\BibitemShut {NoStop}%
\bibitem [{\citenamefont {Morrison}(1958)}]{Morrison1958}%
  \BibitemOpen
  \bibfield  {author} {\bibinfo {author} {\bibfnamefont {P.}~\bibnamefont
  {Morrison}},\ }\bibfield  {title} {\bibinfo {title} {Approximate nature of
  physical symmetries},\ }\href {https://doi.org/10.1119/1.1996159} {\bibfield
  {journal} {\bibinfo  {journal} {American Journal of Physics}\ }\textbf
  {\bibinfo {volume} {26}},\ \bibinfo {pages} {358} (\bibinfo {year}
  {1958})}\BibitemShut {NoStop}%
\bibitem [{\citenamefont {Scherk}(1979)}]{Scherk1979}%
  \BibitemOpen
  \bibfield  {author} {\bibinfo {author} {\bibfnamefont {J.}~\bibnamefont
  {Scherk}},\ }\bibfield  {title} {\bibinfo {title} {Antigravity: A crazy
  idea?},\ }\href {https://doi.org/10.1016/0370-2693(79)90463-5} {\bibfield
  {journal} {\bibinfo  {journal} {Physics Letters B}\ }\textbf {\bibinfo
  {volume} {88}},\ \bibinfo {pages} {265} (\bibinfo {year} {1979})}\BibitemShut
  {NoStop}%
\bibitem [{\citenamefont {Nieto}\ and\ \citenamefont
  {Goldman}(1991)}]{Nieto1991}%
  \BibitemOpen
  \bibfield  {author} {\bibinfo {author} {\bibfnamefont {M.}~\bibnamefont
  {Nieto}}\ and\ \bibinfo {author} {\bibfnamefont {T.}~\bibnamefont
  {Goldman}},\ }\bibfield  {title} {\bibinfo {title} {The arguments against
  antigravity and the gravitational acceleration of antimatter},\ }\href
  {https://doi.org/https://doi.org/10.1016/0370-1573(91)90138-C} {\bibfield
  {journal} {\bibinfo  {journal} {Physics Reports}\ }\textbf {\bibinfo {volume}
  {205}},\ \bibinfo {pages} {221} (\bibinfo {year} {1991})}\BibitemShut
  {NoStop}%
\bibitem [{\citenamefont {Adelberger}\ \emph {et~al.}(1991)\citenamefont
  {Adelberger}, \citenamefont {Heckel}, \citenamefont {Stubbs},\ and\
  \citenamefont {Su}}]{Adelberger1991}%
  \BibitemOpen
  \bibfield  {author} {\bibinfo {author} {\bibfnamefont {E.}~\bibnamefont
  {Adelberger}}, \bibinfo {author} {\bibfnamefont {B.}~\bibnamefont {Heckel}},
  \bibinfo {author} {\bibfnamefont {C.}~\bibnamefont {Stubbs}},\ and\ \bibinfo
  {author} {\bibfnamefont {Y.}~\bibnamefont {Su}},\ }\bibfield  {title}
  {\bibinfo {title} {Does antimatter fall with the same acceleration as
  ordinary matter?},\ }\href {https://doi.org/10.1103/PhysRevLett.66.850}
  {\bibfield  {journal} {\bibinfo  {journal} {Phys. Rev. Lett.}\ }\textbf
  {\bibinfo {volume} {66}},\ \bibinfo {pages} {850} (\bibinfo {year}
  {1991})}\BibitemShut {NoStop}%
\bibitem [{\citenamefont {Huber}\ \emph {et~al.}(2000)\citenamefont {Huber},
  \citenamefont {Lewis}, \citenamefont {Messerschmid},\ and\ \citenamefont
  {Smith}}]{Huber2000}%
  \BibitemOpen
  \bibfield  {author} {\bibinfo {author} {\bibfnamefont {F.}~\bibnamefont
  {Huber}}, \bibinfo {author} {\bibfnamefont {R.}~\bibnamefont {Lewis}},
  \bibinfo {author} {\bibfnamefont {E.}~\bibnamefont {Messerschmid}},\ and\
  \bibinfo {author} {\bibfnamefont {G.}~\bibnamefont {Smith}},\ }\bibfield
  {title} {\bibinfo {title} {Precision tests of einstein’s weak equivalence
  principle for antimatter},\ }\href
  {https://doi.org/10.1016/S0273-1177(99)00999-0} {\bibfield  {journal}
  {\bibinfo  {journal} {Advances in Space Research}\ }\textbf {\bibinfo
  {volume} {25}},\ \bibinfo {pages} {1245} (\bibinfo {year}
  {2000})}\BibitemShut {NoStop}%
\bibitem [{\citenamefont {Chardin}\ and\ \citenamefont
  {Manfredi}(2018)}]{Chardin2018}%
  \BibitemOpen
  \bibfield  {author} {\bibinfo {author} {\bibfnamefont {G.}~\bibnamefont
  {Chardin}}\ and\ \bibinfo {author} {\bibfnamefont {G.}~\bibnamefont
  {Manfredi}},\ }\bibfield  {title} {\bibinfo {title} {Gravity, antimatter and
  the {Dirac-Milne} universe},\ }\href
  {https://doi.org/10.1007/s10751-018-1521-3} {\bibfield  {journal} {\bibinfo
  {journal} {Hyperfine Interactions}\ }\textbf {\bibinfo {volume} {239}},\
  \bibinfo {pages} {45} (\bibinfo {year} {2018})}\BibitemShut {NoStop}%
\bibitem [{\citenamefont {Charman}(2013)}]{Alpha2013}%
  \BibitemOpen
  \bibfield  {author} {\bibinfo {author} {\bibfnamefont {A.~E.}\ \bibnamefont
  {Charman}},\ }\bibfield  {title} {\bibinfo {title} {Description and first
  application of a new technique to measure the gravitational mass of
  antihydrogen},\ }\href {https://doi.org/10.1038/ncomms2787} {\bibfield
  {journal} {\bibinfo  {journal} {Nature Communications}\ ,\ \bibinfo {eid}
  {1785}} (\bibinfo {year} {2013})}\BibitemShut {NoStop}%
\bibitem [{\citenamefont {Borchert}\ \emph {et~al.}(2022)\citenamefont
  {Borchert}, \citenamefont {Devlin}, \citenamefont {Erlewein}, \citenamefont
  {Fleck}, \citenamefont {Harrington}, \citenamefont {Higuchi}, \citenamefont
  {Latacz}, \citenamefont {Voelksen}, \citenamefont {Wursten}, \citenamefont
  {Abbass}, \citenamefont {Bohman}, \citenamefont {Mooser}, \citenamefont
  {Popper}, \citenamefont {Wiesinger}, \citenamefont {Will}, \citenamefont
  {Blaum}, \citenamefont {Matsuda}, \citenamefont {Ospelkaus}, \citenamefont
  {Quint}, \citenamefont {Walz}, \citenamefont {Yamazaki}, \citenamefont
  {Smorra},\ and\ \citenamefont {Ulmer}}]{Borchert2022}%
  \BibitemOpen
  \bibfield  {author} {\bibinfo {author} {\bibfnamefont {M.~J.}\ \bibnamefont
  {Borchert}}, \bibinfo {author} {\bibfnamefont {J.~A.}\ \bibnamefont
  {Devlin}}, \bibinfo {author} {\bibfnamefont {S.~R.}\ \bibnamefont
  {Erlewein}}, \bibinfo {author} {\bibfnamefont {M.}~\bibnamefont {Fleck}},
  \bibinfo {author} {\bibfnamefont {J.~A.}\ \bibnamefont {Harrington}},
  \bibinfo {author} {\bibfnamefont {T.}~\bibnamefont {Higuchi}}, \bibinfo
  {author} {\bibfnamefont {B.~M.}\ \bibnamefont {Latacz}}, \bibinfo {author}
  {\bibfnamefont {F.}~\bibnamefont {Voelksen}}, \bibinfo {author}
  {\bibfnamefont {E.~J.}\ \bibnamefont {Wursten}}, \bibinfo {author}
  {\bibfnamefont {F.}~\bibnamefont {Abbass}}, \bibinfo {author} {\bibfnamefont
  {M.~A.}\ \bibnamefont {Bohman}}, \bibinfo {author} {\bibfnamefont {A.~H.}\
  \bibnamefont {Mooser}}, \bibinfo {author} {\bibfnamefont {D.}~\bibnamefont
  {Popper}}, \bibinfo {author} {\bibfnamefont {M.}~\bibnamefont {Wiesinger}},
  \bibinfo {author} {\bibfnamefont {C.}~\bibnamefont {Will}}, \bibinfo {author}
  {\bibfnamefont {K.}~\bibnamefont {Blaum}}, \bibinfo {author} {\bibfnamefont
  {Y.}~\bibnamefont {Matsuda}}, \bibinfo {author} {\bibfnamefont
  {C.}~\bibnamefont {Ospelkaus}}, \bibinfo {author} {\bibfnamefont
  {W.}~\bibnamefont {Quint}}, \bibinfo {author} {\bibfnamefont
  {J.}~\bibnamefont {Walz}}, \bibinfo {author} {\bibfnamefont {Y.}~\bibnamefont
  {Yamazaki}}, \bibinfo {author} {\bibfnamefont {C.}~\bibnamefont {Smorra}},\
  and\ \bibinfo {author} {\bibfnamefont {S.}~\bibnamefont {Ulmer}},\ }\bibfield
   {title} {\bibinfo {title} {A 16-parts-per-trillion measurement of the
  antiproton-to-proton charge-mass ratio},\ }\href
  {https://doi.org/10.1038/s41586-021-04203-w} {\bibfield  {journal} {\bibinfo
  {journal} {Nature}\ }\textbf {\bibinfo {volume} {601}},\ \bibinfo {pages}
  {53} (\bibinfo {year} {2022})}\BibitemShut {NoStop}%
\bibitem [{\citenamefont {Wagner}\ \emph {et~al.}(2012)\citenamefont {Wagner},
  \citenamefont {Schlamminger}, \citenamefont {Gundlach},\ and\ \citenamefont
  {Adelberger}}]{Wagner2012}%
  \BibitemOpen
  \bibfield  {author} {\bibinfo {author} {\bibfnamefont {T.}~\bibnamefont
  {Wagner}}, \bibinfo {author} {\bibfnamefont {S.}~\bibnamefont
  {Schlamminger}}, \bibinfo {author} {\bibfnamefont {J.}~\bibnamefont
  {Gundlach}},\ and\ \bibinfo {author} {\bibfnamefont {E.}~\bibnamefont
  {Adelberger}},\ }\bibfield  {title} {\bibinfo {title} {Torsion-balance tests
  of the weak equivalence principle},\ }\href
  {https://doi.org/10.1088/0264-9381/29/18/184002} {\bibfield  {journal}
  {\bibinfo  {journal} {Classical and Quantum Gravity}\ }\textbf {\bibinfo
  {volume} {29}},\ \bibinfo {pages} {184002} (\bibinfo {year}
  {2012})}\BibitemShut {NoStop}%
\bibitem [{\citenamefont {Touboul}\ \emph {et~al.}(2017)\citenamefont
  {Touboul}, \citenamefont {M\'etris}, \citenamefont {Rodrigues}, \citenamefont
  {Andr\'e}, \citenamefont {Baghi}, \citenamefont {Berg\'e}, \citenamefont
  {Boulanger}, \citenamefont {Bremer}, \citenamefont {Carle}, \citenamefont
  {Chhun}, \citenamefont {Christophe}, \citenamefont {Cipolla}, \citenamefont
  {Damour}, \citenamefont {Danto}, \citenamefont {Dittus}, \citenamefont
  {Fayet}, \citenamefont {Foulon}, \citenamefont {Gageant}, \citenamefont
  {Guidotti}, \citenamefont {Hagedorn}, \citenamefont {Hardy}, \citenamefont
  {Huynh}, \citenamefont {Inchauspe}, \citenamefont {Kayser}, \citenamefont
  {Lala}, \citenamefont {L\"ammerzahl}, \citenamefont {Lebat}, \citenamefont
  {Leseur}, \citenamefont {Liorzou}, \citenamefont {List}, \citenamefont
  {L\"offler}, \citenamefont {Panet}, \citenamefont {Pouilloux}, \citenamefont
  {Prieur}, \citenamefont {Rebray}, \citenamefont {Reynaud}, \citenamefont
  {Rievers}, \citenamefont {Robert}, \citenamefont {Selig}, \citenamefont
  {Serron}, \citenamefont {Sumner}, \citenamefont {Tanguy},\ and\ \citenamefont
  {Visser}}]{Touboul2017}%
  \BibitemOpen
  \bibfield  {author} {\bibinfo {author} {\bibfnamefont {P.}~\bibnamefont
  {Touboul}}, \bibinfo {author} {\bibfnamefont {G.}~\bibnamefont {M\'etris}},
  \bibinfo {author} {\bibfnamefont {M.}~\bibnamefont {Rodrigues}}, \bibinfo
  {author} {\bibfnamefont {Y.}~\bibnamefont {Andr\'e}}, \bibinfo {author}
  {\bibfnamefont {Q.}~\bibnamefont {Baghi}}, \bibinfo {author} {\bibfnamefont
  {J.}~\bibnamefont {Berg\'e}}, \bibinfo {author} {\bibfnamefont
  {D.}~\bibnamefont {Boulanger}}, \bibinfo {author} {\bibfnamefont
  {S.}~\bibnamefont {Bremer}}, \bibinfo {author} {\bibfnamefont
  {P.}~\bibnamefont {Carle}}, \bibinfo {author} {\bibfnamefont
  {R.}~\bibnamefont {Chhun}}, \bibinfo {author} {\bibfnamefont
  {B.}~\bibnamefont {Christophe}}, \bibinfo {author} {\bibfnamefont
  {V.}~\bibnamefont {Cipolla}}, \bibinfo {author} {\bibfnamefont
  {T.}~\bibnamefont {Damour}}, \bibinfo {author} {\bibfnamefont
  {P.}~\bibnamefont {Danto}}, \bibinfo {author} {\bibfnamefont
  {H.}~\bibnamefont {Dittus}}, \bibinfo {author} {\bibfnamefont
  {P.}~\bibnamefont {Fayet}}, \bibinfo {author} {\bibfnamefont
  {B.}~\bibnamefont {Foulon}}, \bibinfo {author} {\bibfnamefont
  {C.}~\bibnamefont {Gageant}}, \bibinfo {author} {\bibfnamefont {P.-Y.}\
  \bibnamefont {Guidotti}}, \bibinfo {author} {\bibfnamefont {D.}~\bibnamefont
  {Hagedorn}}, \bibinfo {author} {\bibfnamefont {E.}~\bibnamefont {Hardy}},
  \bibinfo {author} {\bibfnamefont {P.-A.}\ \bibnamefont {Huynh}}, \bibinfo
  {author} {\bibfnamefont {H.}~\bibnamefont {Inchauspe}}, \bibinfo {author}
  {\bibfnamefont {P.}~\bibnamefont {Kayser}}, \bibinfo {author} {\bibfnamefont
  {S.}~\bibnamefont {Lala}}, \bibinfo {author} {\bibfnamefont {C.}~\bibnamefont
  {L\"ammerzahl}}, \bibinfo {author} {\bibfnamefont {V.}~\bibnamefont {Lebat}},
  \bibinfo {author} {\bibfnamefont {P.}~\bibnamefont {Leseur}}, \bibinfo
  {author} {\bibfnamefont {F.}~\bibnamefont {Liorzou}}, \bibinfo {author}
  {\bibfnamefont {M.}~\bibnamefont {List}}, \bibinfo {author} {\bibfnamefont
  {F.}~\bibnamefont {L\"offler}}, \bibinfo {author} {\bibfnamefont
  {I.}~\bibnamefont {Panet}}, \bibinfo {author} {\bibfnamefont
  {B.}~\bibnamefont {Pouilloux}}, \bibinfo {author} {\bibfnamefont
  {P.}~\bibnamefont {Prieur}}, \bibinfo {author} {\bibfnamefont
  {A.}~\bibnamefont {Rebray}}, \bibinfo {author} {\bibfnamefont
  {S.}~\bibnamefont {Reynaud}}, \bibinfo {author} {\bibfnamefont
  {B.}~\bibnamefont {Rievers}}, \bibinfo {author} {\bibfnamefont
  {A.}~\bibnamefont {Robert}}, \bibinfo {author} {\bibfnamefont
  {H.}~\bibnamefont {Selig}}, \bibinfo {author} {\bibfnamefont
  {L.}~\bibnamefont {Serron}}, \bibinfo {author} {\bibfnamefont
  {T.}~\bibnamefont {Sumner}}, \bibinfo {author} {\bibfnamefont
  {N.}~\bibnamefont {Tanguy}},\ and\ \bibinfo {author} {\bibfnamefont
  {P.}~\bibnamefont {Visser}},\ }\bibfield  {title} {\bibinfo {title}
  {{MICROSCOPE} mission: First results of a space test of the equivalence
  principle},\ }\href {https://doi.org/10.1103/PhysRevLett.119.231101}
  {\bibfield  {journal} {\bibinfo  {journal} {Phys. Rev. Lett.}\ }\textbf
  {\bibinfo {volume} {119}},\ \bibinfo {pages} {231101} (\bibinfo {year}
  {2017})}\BibitemShut {NoStop}%
\bibitem [{\citenamefont {Will}(2018)}]{Will2018}%
  \BibitemOpen
  \bibfield  {author} {\bibinfo {author} {\bibfnamefont {C.}~\bibnamefont
  {Will}},\ }\href@noop {} {\emph {\bibinfo {title} {Theory and Experiment in
  Gravitational Physics (new edition)}}}\ (\bibinfo  {publisher} {{Cambridge
  University Press}},\ \bibinfo {year} {2018})\BibitemShut {NoStop}%
\bibitem [{\citenamefont {Viswanathan}\ \emph {et~al.}(2018)\citenamefont
  {Viswanathan}, \citenamefont {Fienga}, \citenamefont {Minazzoli},
  \citenamefont {Bernus}, \citenamefont {Laskar},\ and\ \citenamefont
  {Gastineau}}]{Viswanathan2018}%
  \BibitemOpen
  \bibfield  {author} {\bibinfo {author} {\bibfnamefont {V.}~\bibnamefont
  {Viswanathan}}, \bibinfo {author} {\bibfnamefont {A.}~\bibnamefont {Fienga}},
  \bibinfo {author} {\bibfnamefont {O.}~\bibnamefont {Minazzoli}}, \bibinfo
  {author} {\bibfnamefont {L.}~\bibnamefont {Bernus}}, \bibinfo {author}
  {\bibfnamefont {J.}~\bibnamefont {Laskar}},\ and\ \bibinfo {author}
  {\bibfnamefont {M.}~\bibnamefont {Gastineau}},\ }\bibfield  {title} {\bibinfo
  {title} {The new lunar ephemeris {INPOP17a} and its application to
  fundamental physics},\ }\href {https://doi.org/10.1093/mnras/sty096}
  {\bibfield  {journal} {\bibinfo  {journal} {Monthly Notices of the Royal
  Astronomical Society}\ }\textbf {\bibinfo {volume} {476}},\ \bibinfo {pages}
  {1877} (\bibinfo {year} {2018})}\BibitemShut {NoStop}%
\bibitem [{\citenamefont {Maury}\ \emph {et~al.}(2014)\citenamefont {Maury},
  \citenamefont {Oelert}, \citenamefont {Bartmann}, \citenamefont
  {Belochitskii}, \citenamefont {Breuker}, \citenamefont {Butin}, \citenamefont
  {Carli}, \citenamefont {Eriksson}, \citenamefont {Pasinelli},\ and\
  \citenamefont {Tranquille}}]{Maury2014}%
  \BibitemOpen
  \bibfield  {author} {\bibinfo {author} {\bibfnamefont {S.}~\bibnamefont
  {Maury}}, \bibinfo {author} {\bibfnamefont {W.}~\bibnamefont {Oelert}},
  \bibinfo {author} {\bibfnamefont {W.}~\bibnamefont {Bartmann}}, \bibinfo
  {author} {\bibfnamefont {P.}~\bibnamefont {Belochitskii}}, \bibinfo {author}
  {\bibfnamefont {H.}~\bibnamefont {Breuker}}, \bibinfo {author} {\bibfnamefont
  {F.}~\bibnamefont {Butin}}, \bibinfo {author} {\bibfnamefont
  {C.}~\bibnamefont {Carli}}, \bibinfo {author} {\bibfnamefont
  {T.}~\bibnamefont {Eriksson}}, \bibinfo {author} {\bibfnamefont
  {S.}~\bibnamefont {Pasinelli}},\ and\ \bibinfo {author} {\bibfnamefont
  {G.}~\bibnamefont {Tranquille}},\ }\bibfield  {title} {\bibinfo {title}
  {{{ELENA}}: the extra low energy anti-proton facility at {{CERN}}},\ }\href
  {https://doi.org/10.1007/s10751-014-1067-y} {\bibfield  {journal} {\bibinfo
  {journal} {Hyperfine Interactions}\ }\textbf {\bibinfo {volume} {229}},\
  \bibinfo {pages} {105} (\bibinfo {year} {2014})}\BibitemShut {NoStop}%
\bibitem [{\citenamefont {Indelicato}\ \emph {et~al.}(2014)\citenamefont
  {Indelicato}, \citenamefont {Chardin}, \citenamefont {Grandemange},
  \citenamefont {Lunney}, \citenamefont {Manea}, \citenamefont {Badertscher},
  \citenamefont {Crivelli}, \citenamefont {Curioni}, \citenamefont
  {Marchionni}, \citenamefont {Rossi}, \citenamefont {Rubbia}, \citenamefont
  {Nesvizhevsky}, \citenamefont {{Brook-Roberge}}, \citenamefont {Comini},
  \citenamefont {Debu}, \citenamefont {Dupr{\'e}}, \citenamefont {Liszkay},
  \citenamefont {Mansouli{\'e}}, \citenamefont {P{\'e}rez}, \citenamefont
  {Rey}, \citenamefont {Reymond}, \citenamefont {Ruiz}, \citenamefont
  {Sacquin}, \citenamefont {Vallage}, \citenamefont {Biraben}, \citenamefont
  {Clad{\'e}}, \citenamefont {Douillet}, \citenamefont {Dufour}, \citenamefont
  {Guellati}, \citenamefont {Hilico}, \citenamefont {Lambrecht}, \citenamefont
  {Gu{\'e}rout}, \citenamefont {Karr}, \citenamefont {Nez}, \citenamefont
  {Reynaud}, \citenamefont {Szabo}, \citenamefont {Tran}, \citenamefont
  {Trapateau}, \citenamefont {Mohri}, \citenamefont {Yamazaki}, \citenamefont
  {Charlton}, \citenamefont {Eriksson}, \citenamefont {Madsen}, \citenamefont
  {Werf}, \citenamefont {Kuroda}, \citenamefont {Torii}, \citenamefont
  {Nagashima}, \citenamefont {{Schmidt-Kaler}}, \citenamefont {Walz},
  \citenamefont {Wolf}, \citenamefont {Hervieux}, \citenamefont {Manfredi},
  \citenamefont {Voronin}, \citenamefont {Froelich}, \citenamefont {Wronka},\
  and\ \citenamefont {Staszczak}}]{Indelicato2014}%
  \BibitemOpen
  \bibfield  {author} {\bibinfo {author} {\bibfnamefont {P.}~\bibnamefont
  {Indelicato}}, \bibinfo {author} {\bibfnamefont {G.}~\bibnamefont {Chardin}},
  \bibinfo {author} {\bibfnamefont {P.}~\bibnamefont {Grandemange}}, \bibinfo
  {author} {\bibfnamefont {D.}~\bibnamefont {Lunney}}, \bibinfo {author}
  {\bibfnamefont {V.}~\bibnamefont {Manea}}, \bibinfo {author} {\bibfnamefont
  {A.}~\bibnamefont {Badertscher}}, \bibinfo {author} {\bibfnamefont
  {P.}~\bibnamefont {Crivelli}}, \bibinfo {author} {\bibfnamefont
  {A.}~\bibnamefont {Curioni}}, \bibinfo {author} {\bibfnamefont
  {A.}~\bibnamefont {Marchionni}}, \bibinfo {author} {\bibfnamefont
  {B.}~\bibnamefont {Rossi}}, \bibinfo {author} {\bibfnamefont
  {A.}~\bibnamefont {Rubbia}}, \bibinfo {author} {\bibfnamefont
  {V.}~\bibnamefont {Nesvizhevsky}}, \bibinfo {author} {\bibfnamefont
  {D.}~\bibnamefont {{Brook-Roberge}}}, \bibinfo {author} {\bibfnamefont
  {P.}~\bibnamefont {Comini}}, \bibinfo {author} {\bibfnamefont
  {P.}~\bibnamefont {Debu}}, \bibinfo {author} {\bibfnamefont {P.}~\bibnamefont
  {Dupr{\'e}}}, \bibinfo {author} {\bibfnamefont {L.}~\bibnamefont {Liszkay}},
  \bibinfo {author} {\bibfnamefont {B.}~\bibnamefont {Mansouli{\'e}}}, \bibinfo
  {author} {\bibfnamefont {P.}~\bibnamefont {P{\'e}rez}}, \bibinfo {author}
  {\bibfnamefont {J.-M.}\ \bibnamefont {Rey}}, \bibinfo {author} {\bibfnamefont
  {B.}~\bibnamefont {Reymond}}, \bibinfo {author} {\bibfnamefont
  {N.}~\bibnamefont {Ruiz}}, \bibinfo {author} {\bibfnamefont {Y.}~\bibnamefont
  {Sacquin}}, \bibinfo {author} {\bibfnamefont {B.}~\bibnamefont {Vallage}},
  \bibinfo {author} {\bibfnamefont {F.}~\bibnamefont {Biraben}}, \bibinfo
  {author} {\bibfnamefont {P.}~\bibnamefont {Clad{\'e}}}, \bibinfo {author}
  {\bibfnamefont {A.}~\bibnamefont {Douillet}}, \bibinfo {author}
  {\bibfnamefont {G.}~\bibnamefont {Dufour}}, \bibinfo {author} {\bibfnamefont
  {S.}~\bibnamefont {Guellati}}, \bibinfo {author} {\bibfnamefont
  {L.}~\bibnamefont {Hilico}}, \bibinfo {author} {\bibfnamefont
  {A.}~\bibnamefont {Lambrecht}}, \bibinfo {author} {\bibfnamefont
  {R.}~\bibnamefont {Gu{\'e}rout}}, \bibinfo {author} {\bibfnamefont {J.-P.}\
  \bibnamefont {Karr}}, \bibinfo {author} {\bibfnamefont {F.}~\bibnamefont
  {Nez}}, \bibinfo {author} {\bibfnamefont {S.}~\bibnamefont {Reynaud}},
  \bibinfo {author} {\bibfnamefont {I.~C.}\ \bibnamefont {Szabo}}, \bibinfo
  {author} {\bibfnamefont {V.-Q.}\ \bibnamefont {Tran}}, \bibinfo {author}
  {\bibfnamefont {J.}~\bibnamefont {Trapateau}}, \bibinfo {author}
  {\bibfnamefont {A.}~\bibnamefont {Mohri}}, \bibinfo {author} {\bibfnamefont
  {Y.}~\bibnamefont {Yamazaki}}, \bibinfo {author} {\bibfnamefont
  {M.}~\bibnamefont {Charlton}}, \bibinfo {author} {\bibfnamefont
  {S.}~\bibnamefont {Eriksson}}, \bibinfo {author} {\bibfnamefont
  {N.}~\bibnamefont {Madsen}}, \bibinfo {author} {\bibfnamefont
  {D.}~\bibnamefont {Werf}}, \bibinfo {author} {\bibfnamefont {N.}~\bibnamefont
  {Kuroda}}, \bibinfo {author} {\bibfnamefont {H.}~\bibnamefont {Torii}},
  \bibinfo {author} {\bibfnamefont {Y.}~\bibnamefont {Nagashima}}, \bibinfo
  {author} {\bibfnamefont {F.}~\bibnamefont {{Schmidt-Kaler}}}, \bibinfo
  {author} {\bibfnamefont {J.}~\bibnamefont {Walz}}, \bibinfo {author}
  {\bibfnamefont {S.}~\bibnamefont {Wolf}}, \bibinfo {author} {\bibfnamefont
  {P.-A.}\ \bibnamefont {Hervieux}}, \bibinfo {author} {\bibfnamefont
  {G.}~\bibnamefont {Manfredi}}, \bibinfo {author} {\bibfnamefont
  {A.}~\bibnamefont {Voronin}}, \bibinfo {author} {\bibfnamefont
  {P.}~\bibnamefont {Froelich}}, \bibinfo {author} {\bibfnamefont
  {S.}~\bibnamefont {Wronka}},\ and\ \bibinfo {author} {\bibfnamefont
  {M.}~\bibnamefont {Staszczak}},\ }\bibfield  {title} {\bibinfo {title} {The
  {{GBAR}} project, or how does antimatter fall?},\ }\href
  {https://doi.org/10.1007/s10751-014-1019-6} {\bibfield  {journal} {\bibinfo
  {journal} {Hyperfine Interactions}\ }\textbf {\bibinfo {volume} {228}},\
  \bibinfo {pages} {141} (\bibinfo {year} {2014})}\BibitemShut {NoStop}%
\bibitem [{\citenamefont {Bertsche}(2018)}]{Bertsche2018}%
  \BibitemOpen
  \bibfield  {author} {\bibinfo {author} {\bibfnamefont {W.}~\bibnamefont
  {Bertsche}},\ }\bibfield  {title} {\bibinfo {title} {Prospects for comparison
  of matter and antimatter gravitation with {{ALPHA}}-g},\ }\href
  {https://doi.org/10.1098/rsta.2017.0265} {\bibfield  {journal} {\bibinfo
  {journal} {Philosophical Transactions of the Royal Society A: Mathematical,
  Physical and Engineering Sciences}\ }\textbf {\bibinfo {volume} {376}},\
  \bibinfo {pages} {20170265} (\bibinfo {year} {2018})}\BibitemShut {NoStop}%
\bibitem [{\citenamefont {Pagano}\ \emph {et~al.}(2020)\citenamefont {Pagano},
  \citenamefont {Aghion}, \citenamefont {Amsler}, \citenamefont {Bonomi},
  \citenamefont {Brusa}, \citenamefont {Caccia}, \citenamefont {Caravita},
  \citenamefont {Castelli}, \citenamefont {Cerchiari}, \citenamefont
  {Comparat}, \citenamefont {Consolati}, \citenamefont {Demetrio},
  \citenamefont {Noto}, \citenamefont {Doser}, \citenamefont {Evans},
  \citenamefont {Fani}, \citenamefont {Ferragut}, \citenamefont {Fesel},
  \citenamefont {Fontana}, \citenamefont {Gerber}, \citenamefont {Giammarchi},
  \citenamefont {Gligorova}, \citenamefont {Guatieri}, \citenamefont {Haider},
  \citenamefont {Hinterberger}, \citenamefont {Holmestad}, \citenamefont
  {Kellerbauer}, \citenamefont {Khalidova}, \citenamefont {Krasnick{\'{y}}},
  \citenamefont {Lagomarsino}, \citenamefont {Lansonneur}, \citenamefont
  {Lebrun}, \citenamefont {Malbrunot}, \citenamefont {Mariazzi}, \citenamefont
  {Marton}, \citenamefont {Matveev}, \citenamefont {Mazzotta}, \citenamefont
  {MÃ¼ller}, \citenamefont {Nebbia}, \citenamefont {Nedelec}, \citenamefont
  {Oberthaler}, \citenamefont {Pacifico}, \citenamefont {Penasa}, \citenamefont
  {Petracek}, \citenamefont {Prelz}, \citenamefont {Prevedelli}, \citenamefont
  {Ravelli}, \citenamefont {Rienaecker}, \citenamefont {Robert}, \citenamefont
  {R{\o}hne}, \citenamefont {Rotondi}, \citenamefont {Sandaker}, \citenamefont
  {Santoro}, \citenamefont {Smestad}, \citenamefont {Sorrentino}, \citenamefont
  {Testera}, \citenamefont {Tietje}, \citenamefont {Widmann}, \citenamefont
  {Yzombard}, \citenamefont {Zimmer}, \citenamefont {Zmeskal},\ and\
  \citenamefont {Zurlo}}]{Pagano2020}%
  \BibitemOpen
  \bibfield  {author} {\bibinfo {author} {\bibfnamefont {D.}~\bibnamefont
  {Pagano}}, \bibinfo {author} {\bibfnamefont {S.}~\bibnamefont {Aghion}},
  \bibinfo {author} {\bibfnamefont {C.}~\bibnamefont {Amsler}}, \bibinfo
  {author} {\bibfnamefont {G.}~\bibnamefont {Bonomi}}, \bibinfo {author}
  {\bibfnamefont {R.~S.}\ \bibnamefont {Brusa}}, \bibinfo {author}
  {\bibfnamefont {M.}~\bibnamefont {Caccia}}, \bibinfo {author} {\bibfnamefont
  {R.}~\bibnamefont {Caravita}}, \bibinfo {author} {\bibfnamefont
  {F.}~\bibnamefont {Castelli}}, \bibinfo {author} {\bibfnamefont
  {G.}~\bibnamefont {Cerchiari}}, \bibinfo {author} {\bibfnamefont
  {D.}~\bibnamefont {Comparat}}, \bibinfo {author} {\bibfnamefont
  {G.}~\bibnamefont {Consolati}}, \bibinfo {author} {\bibfnamefont
  {A.}~\bibnamefont {Demetrio}}, \bibinfo {author} {\bibfnamefont
  {L.}~\bibnamefont {Noto}}, \bibinfo {author} {\bibfnamefont {M.}~\bibnamefont
  {Doser}}, \bibinfo {author} {\bibfnamefont {A.}~\bibnamefont {Evans}},
  \bibinfo {author} {\bibfnamefont {M.}~\bibnamefont {Fani}}, \bibinfo {author}
  {\bibfnamefont {R.}~\bibnamefont {Ferragut}}, \bibinfo {author}
  {\bibfnamefont {J.}~\bibnamefont {Fesel}}, \bibinfo {author} {\bibfnamefont
  {A.}~\bibnamefont {Fontana}}, \bibinfo {author} {\bibfnamefont
  {S.}~\bibnamefont {Gerber}}, \bibinfo {author} {\bibfnamefont
  {M.}~\bibnamefont {Giammarchi}}, \bibinfo {author} {\bibfnamefont
  {A.}~\bibnamefont {Gligorova}}, \bibinfo {author} {\bibfnamefont
  {F.}~\bibnamefont {Guatieri}}, \bibinfo {author} {\bibfnamefont
  {S.}~\bibnamefont {Haider}}, \bibinfo {author} {\bibfnamefont
  {A.}~\bibnamefont {Hinterberger}}, \bibinfo {author} {\bibfnamefont
  {H.}~\bibnamefont {Holmestad}}, \bibinfo {author} {\bibfnamefont
  {A.}~\bibnamefont {Kellerbauer}}, \bibinfo {author} {\bibfnamefont
  {O.}~\bibnamefont {Khalidova}}, \bibinfo {author} {\bibfnamefont
  {D.}~\bibnamefont {Krasnick{\'{y}}}}, \bibinfo {author} {\bibfnamefont
  {V.}~\bibnamefont {Lagomarsino}}, \bibinfo {author} {\bibfnamefont
  {P.}~\bibnamefont {Lansonneur}}, \bibinfo {author} {\bibfnamefont
  {P.}~\bibnamefont {Lebrun}}, \bibinfo {author} {\bibfnamefont
  {C.}~\bibnamefont {Malbrunot}}, \bibinfo {author} {\bibfnamefont
  {S.}~\bibnamefont {Mariazzi}}, \bibinfo {author} {\bibfnamefont
  {J.}~\bibnamefont {Marton}}, \bibinfo {author} {\bibfnamefont
  {V.}~\bibnamefont {Matveev}}, \bibinfo {author} {\bibfnamefont
  {Z.}~\bibnamefont {Mazzotta}}, \bibinfo {author} {\bibfnamefont {S.~R.}\
  \bibnamefont {MÃ¼ller}}, \bibinfo {author} {\bibfnamefont {G.}~\bibnamefont
  {Nebbia}}, \bibinfo {author} {\bibfnamefont {P.}~\bibnamefont {Nedelec}},
  \bibinfo {author} {\bibfnamefont {M.}~\bibnamefont {Oberthaler}}, \bibinfo
  {author} {\bibfnamefont {N.}~\bibnamefont {Pacifico}}, \bibinfo {author}
  {\bibfnamefont {L.}~\bibnamefont {Penasa}}, \bibinfo {author} {\bibfnamefont
  {V.}~\bibnamefont {Petracek}}, \bibinfo {author} {\bibfnamefont
  {F.}~\bibnamefont {Prelz}}, \bibinfo {author} {\bibfnamefont
  {M.}~\bibnamefont {Prevedelli}}, \bibinfo {author} {\bibfnamefont
  {L.}~\bibnamefont {Ravelli}}, \bibinfo {author} {\bibfnamefont
  {B.}~\bibnamefont {Rienaecker}}, \bibinfo {author} {\bibfnamefont
  {J.}~\bibnamefont {Robert}}, \bibinfo {author} {\bibfnamefont
  {O.}~\bibnamefont {R{\o}hne}}, \bibinfo {author} {\bibfnamefont
  {A.}~\bibnamefont {Rotondi}}, \bibinfo {author} {\bibfnamefont
  {H.}~\bibnamefont {Sandaker}}, \bibinfo {author} {\bibfnamefont
  {R.}~\bibnamefont {Santoro}}, \bibinfo {author} {\bibfnamefont
  {L.}~\bibnamefont {Smestad}}, \bibinfo {author} {\bibfnamefont
  {F.}~\bibnamefont {Sorrentino}}, \bibinfo {author} {\bibfnamefont
  {G.}~\bibnamefont {Testera}}, \bibinfo {author} {\bibfnamefont {I.~C.}\
  \bibnamefont {Tietje}}, \bibinfo {author} {\bibfnamefont {E.}~\bibnamefont
  {Widmann}}, \bibinfo {author} {\bibfnamefont {P.}~\bibnamefont {Yzombard}},
  \bibinfo {author} {\bibfnamefont {C.}~\bibnamefont {Zimmer}}, \bibinfo
  {author} {\bibfnamefont {J.}~\bibnamefont {Zmeskal}},\ and\ \bibinfo {author}
  {\bibfnamefont {N.}~\bibnamefont {Zurlo}},\ }\bibfield  {title} {\bibinfo
  {title} {Gravity and antimatter: the {AEgIS} experiment at {CERN}},\ }\href
  {https://doi.org/10.1088/1742-6596/1342/1/012016} {\bibfield  {journal}
  {\bibinfo  {journal} {Journal of Physics: Conference Series}\ }\textbf
  {\bibinfo {volume} {1342}},\ \bibinfo {pages} {012016} (\bibinfo {year}
  {2020})}\BibitemShut {NoStop}%
\bibitem [{\citenamefont {P{\'e}rez}\ \emph {et~al.}(2015)\citenamefont
  {P{\'e}rez}, \citenamefont {Banerjee}, \citenamefont {Biraben}, \citenamefont
  {{Brook-Roberge}}, \citenamefont {Charlton}, \citenamefont {Clad{\'e}},
  \citenamefont {Comini}, \citenamefont {Crivelli}, \citenamefont {Dalkarov},
  \citenamefont {Debu}, \citenamefont {Douillet}, \citenamefont {Dufour},
  \citenamefont {Dupr{\'e}}, \citenamefont {Eriksson}, \citenamefont
  {Froelich}, \citenamefont {Grandemange}, \citenamefont {Guellati},
  \citenamefont {Gu{\'e}rout}, \citenamefont {Heinrich}, \citenamefont
  {Hervieux}, \citenamefont {Hilico}, \citenamefont {Husson}, \citenamefont
  {Indelicato}, \citenamefont {Jonsell}, \citenamefont {Karr}, \citenamefont
  {Khabarova}, \citenamefont {Kolachevsky}, \citenamefont {Kuroda},
  \citenamefont {Lambrecht}, \citenamefont {Leite}, \citenamefont {Liszkay},
  \citenamefont {Lunney}, \citenamefont {Madsen}, \citenamefont {Manfredi},
  \citenamefont {Mansouli{\'e}}, \citenamefont {Matsuda}, \citenamefont
  {Mohri}, \citenamefont {Mortensen}, \citenamefont {Nagashima}, \citenamefont
  {Nesvizhevsky}, \citenamefont {Nez}, \citenamefont {Regenfus}, \citenamefont
  {Rey}, \citenamefont {Reymond}, \citenamefont {Reynaud}, \citenamefont
  {Rubbia}, \citenamefont {Sacquin}, \citenamefont {{Schmidt-Kaler}},
  \citenamefont {Sillitoe}, \citenamefont {Staszczak}, \citenamefont
  {{Szabo-Foster}}, \citenamefont {Torii}, \citenamefont {Vallage},
  \citenamefont {Valdes}, \citenamefont {{Van der Werf}}, \citenamefont
  {Voronin}, \citenamefont {Walz}, \citenamefont {Wolf}, \citenamefont
  {Wronka},\ and\ \citenamefont {Yamazaki}}]{Perez2015}%
  \BibitemOpen
  \bibfield  {author} {\bibinfo {author} {\bibfnamefont {P.}~\bibnamefont
  {P{\'e}rez}}, \bibinfo {author} {\bibfnamefont {D.}~\bibnamefont {Banerjee}},
  \bibinfo {author} {\bibfnamefont {F.}~\bibnamefont {Biraben}}, \bibinfo
  {author} {\bibfnamefont {D.}~\bibnamefont {{Brook-Roberge}}}, \bibinfo
  {author} {\bibfnamefont {M.}~\bibnamefont {Charlton}}, \bibinfo {author}
  {\bibfnamefont {P.}~\bibnamefont {Clad{\'e}}}, \bibinfo {author}
  {\bibfnamefont {P.}~\bibnamefont {Comini}}, \bibinfo {author} {\bibfnamefont
  {P.}~\bibnamefont {Crivelli}}, \bibinfo {author} {\bibfnamefont
  {O.}~\bibnamefont {Dalkarov}}, \bibinfo {author} {\bibfnamefont
  {P.}~\bibnamefont {Debu}}, \bibinfo {author} {\bibfnamefont {A.}~\bibnamefont
  {Douillet}}, \bibinfo {author} {\bibfnamefont {G.}~\bibnamefont {Dufour}},
  \bibinfo {author} {\bibfnamefont {P.}~\bibnamefont {Dupr{\'e}}}, \bibinfo
  {author} {\bibfnamefont {S.}~\bibnamefont {Eriksson}}, \bibinfo {author}
  {\bibfnamefont {P.}~\bibnamefont {Froelich}}, \bibinfo {author}
  {\bibfnamefont {P.}~\bibnamefont {Grandemange}}, \bibinfo {author}
  {\bibfnamefont {S.}~\bibnamefont {Guellati}}, \bibinfo {author}
  {\bibfnamefont {R.}~\bibnamefont {Gu{\'e}rout}}, \bibinfo {author}
  {\bibfnamefont {J.~M.}\ \bibnamefont {Heinrich}}, \bibinfo {author}
  {\bibfnamefont {P.-A.}\ \bibnamefont {Hervieux}}, \bibinfo {author}
  {\bibfnamefont {L.}~\bibnamefont {Hilico}}, \bibinfo {author} {\bibfnamefont
  {A.}~\bibnamefont {Husson}}, \bibinfo {author} {\bibfnamefont
  {P.}~\bibnamefont {Indelicato}}, \bibinfo {author} {\bibfnamefont
  {S.}~\bibnamefont {Jonsell}}, \bibinfo {author} {\bibfnamefont {J.-P.}\
  \bibnamefont {Karr}}, \bibinfo {author} {\bibfnamefont {K.}~\bibnamefont
  {Khabarova}}, \bibinfo {author} {\bibfnamefont {N.}~\bibnamefont
  {Kolachevsky}}, \bibinfo {author} {\bibfnamefont {N.}~\bibnamefont {Kuroda}},
  \bibinfo {author} {\bibfnamefont {A.}~\bibnamefont {Lambrecht}}, \bibinfo
  {author} {\bibfnamefont {A.~M.~M.}\ \bibnamefont {Leite}}, \bibinfo {author}
  {\bibfnamefont {L.}~\bibnamefont {Liszkay}}, \bibinfo {author} {\bibfnamefont
  {D.}~\bibnamefont {Lunney}}, \bibinfo {author} {\bibfnamefont
  {N.}~\bibnamefont {Madsen}}, \bibinfo {author} {\bibfnamefont
  {G.}~\bibnamefont {Manfredi}}, \bibinfo {author} {\bibfnamefont
  {B.}~\bibnamefont {Mansouli{\'e}}}, \bibinfo {author} {\bibfnamefont
  {Y.}~\bibnamefont {Matsuda}}, \bibinfo {author} {\bibfnamefont
  {A.}~\bibnamefont {Mohri}}, \bibinfo {author} {\bibfnamefont
  {T.}~\bibnamefont {Mortensen}}, \bibinfo {author} {\bibfnamefont
  {Y.}~\bibnamefont {Nagashima}}, \bibinfo {author} {\bibfnamefont
  {V.}~\bibnamefont {Nesvizhevsky}}, \bibinfo {author} {\bibfnamefont
  {F.}~\bibnamefont {Nez}}, \bibinfo {author} {\bibfnamefont {C.}~\bibnamefont
  {Regenfus}}, \bibinfo {author} {\bibfnamefont {J.-M.}\ \bibnamefont {Rey}},
  \bibinfo {author} {\bibfnamefont {J.-M.}\ \bibnamefont {Reymond}}, \bibinfo
  {author} {\bibfnamefont {S.}~\bibnamefont {Reynaud}}, \bibinfo {author}
  {\bibfnamefont {A.}~\bibnamefont {Rubbia}}, \bibinfo {author} {\bibfnamefont
  {Y.}~\bibnamefont {Sacquin}}, \bibinfo {author} {\bibfnamefont
  {F.}~\bibnamefont {{Schmidt-Kaler}}}, \bibinfo {author} {\bibfnamefont
  {N.}~\bibnamefont {Sillitoe}}, \bibinfo {author} {\bibfnamefont
  {M.}~\bibnamefont {Staszczak}}, \bibinfo {author} {\bibfnamefont {C.~I.}\
  \bibnamefont {{Szabo-Foster}}}, \bibinfo {author} {\bibfnamefont
  {H.}~\bibnamefont {Torii}}, \bibinfo {author} {\bibfnamefont
  {B.}~\bibnamefont {Vallage}}, \bibinfo {author} {\bibfnamefont
  {M.}~\bibnamefont {Valdes}}, \bibinfo {author} {\bibfnamefont {D.~P.}\
  \bibnamefont {{Van der Werf}}}, \bibinfo {author} {\bibfnamefont
  {A.}~\bibnamefont {Voronin}}, \bibinfo {author} {\bibfnamefont
  {J.}~\bibnamefont {Walz}}, \bibinfo {author} {\bibfnamefont {S.}~\bibnamefont
  {Wolf}}, \bibinfo {author} {\bibfnamefont {S.}~\bibnamefont {Wronka}},\ and\
  \bibinfo {author} {\bibfnamefont {Y.}~\bibnamefont {Yamazaki}},\ }\bibfield
  {title} {\bibinfo {title} {The {{GBAR}} antimatter gravity experiment},\
  }\href {https://doi.org/10.1007/s10751-015-1154-8} {\bibfield  {journal}
  {\bibinfo  {journal} {Hyperfine Interactions}\ }\textbf {\bibinfo {volume}
  {233}},\ \bibinfo {pages} {21} (\bibinfo {year} {2015})}\BibitemShut
  {NoStop}%
\bibitem [{\citenamefont {Walz}\ and\ \citenamefont
  {H{\"a}nsch}(2004)}]{Walz2004}%
  \BibitemOpen
  \bibfield  {author} {\bibinfo {author} {\bibfnamefont {J.}~\bibnamefont
  {Walz}}\ and\ \bibinfo {author} {\bibfnamefont {T.}~\bibnamefont
  {H{\"a}nsch}},\ }\bibfield  {title} {\bibinfo {title} {A proposal to measure
  antimatter gravity using ultracold antihydrogen atoms},\ }\href
  {https://doi.org/10.1023/B:GERG.0000010730.93408.87} {\bibfield  {journal}
  {\bibinfo  {journal} {General Relativity and Gravitation}\ }\textbf {\bibinfo
  {volume} {36}},\ \bibinfo {pages} {561} (\bibinfo {year} {2004})}\BibitemShut
  {NoStop}%
\bibitem [{\citenamefont {Hilico}\ \emph {et~al.}(2014)\citenamefont {Hilico},
  \citenamefont {Karr}, \citenamefont {Douillet}, \citenamefont {Indelicato},
  \citenamefont {Wolf},\ and\ \citenamefont {Schmidt-Kaler}}]{Hilico2014}%
  \BibitemOpen
  \bibfield  {author} {\bibinfo {author} {\bibfnamefont {L.}~\bibnamefont
  {Hilico}}, \bibinfo {author} {\bibfnamefont {J.-P.}\ \bibnamefont {Karr}},
  \bibinfo {author} {\bibfnamefont {A.}~\bibnamefont {Douillet}}, \bibinfo
  {author} {\bibfnamefont {P.}~\bibnamefont {Indelicato}}, \bibinfo {author}
  {\bibfnamefont {S.}~\bibnamefont {Wolf}},\ and\ \bibinfo {author}
  {\bibfnamefont {F.}~\bibnamefont {Schmidt-Kaler}},\ }\bibfield  {title}
  {\bibinfo {title} {Preparing single ultra-cold antihydrogen atoms for
  free-fall in {GBAR}},\ }\href {https://doi.org/10.1142/S2010194514602695}
  {\bibfield  {journal} {\bibinfo  {journal} {International Journal of Modern
  Physics: Conference Series}\ }\textbf {\bibinfo {volume} {30}},\ \bibinfo
  {pages} {1460269} (\bibinfo {year} {2014})}\BibitemShut {NoStop}%
\bibitem [{\citenamefont {Sillitoe}\ \emph {et~al.}(2017)\citenamefont
  {Sillitoe}, \citenamefont {Karr}, \citenamefont {Heinrich}, \citenamefont
  {Louvradoux}, \citenamefont {Douillet},\ and\ \citenamefont
  {Hilico}}]{Sillitoe2017}%
  \BibitemOpen
  \bibfield  {author} {\bibinfo {author} {\bibfnamefont {N.}~\bibnamefont
  {Sillitoe}}, \bibinfo {author} {\bibfnamefont {J.-P.}\ \bibnamefont {Karr}},
  \bibinfo {author} {\bibfnamefont {J.}~\bibnamefont {Heinrich}}, \bibinfo
  {author} {\bibfnamefont {T.}~\bibnamefont {Louvradoux}}, \bibinfo {author}
  {\bibfnamefont {A.}~\bibnamefont {Douillet}},\ and\ \bibinfo {author}
  {\bibfnamefont {L.}~\bibnamefont {Hilico}},\ }\bibfield  {title} {\bibinfo
  {title} {{$\overline{\mathrm{H}}^{+}$ Sympathetic Cooling Simulations with a
  Variable Time Step}},\ }\href {https://doi.org/10.7566/JPSCP.18.011014}
  {\bibfield  {journal} {\bibinfo  {journal} {JPS Conf. Proc.}\ }\textbf
  {\bibinfo {volume} {18}},\ \bibinfo {pages} {011014} (\bibinfo {year}
  {2017})}\BibitemShut {NoStop}%
\bibitem [{\citenamefont {Lykke}\ \emph {et~al.}(1991)\citenamefont {Lykke},
  \citenamefont {Murray},\ and\ \citenamefont {Lineberger}}]{Lykke1991}%
  \BibitemOpen
  \bibfield  {author} {\bibinfo {author} {\bibfnamefont {K.}~\bibnamefont
  {Lykke}}, \bibinfo {author} {\bibfnamefont {K.}~\bibnamefont {Murray}},\ and\
  \bibinfo {author} {\bibfnamefont {W.}~\bibnamefont {Lineberger}},\ }\bibfield
   {title} {\bibinfo {title} {Threshold photodetachment of
  {{${\mathrm{H}}^{\mathrm{\ensuremath{-}}}$}}},\ }\href
  {https://doi.org/10.1103/PhysRevA.43.6104} {\bibfield  {journal} {\bibinfo
  {journal} {Physical Review A}\ }\textbf {\bibinfo {volume} {43}},\ \bibinfo
  {pages} {6104} (\bibinfo {year} {1991})}\BibitemShut {NoStop}%
\bibitem [{\citenamefont {Vandevraye}\ \emph {et~al.}(2014)\citenamefont
  {Vandevraye}, \citenamefont {Babilotte}, \citenamefont {Drag},\ and\
  \citenamefont {Blondel}}]{Vandevraye2014}%
  \BibitemOpen
  \bibfield  {author} {\bibinfo {author} {\bibfnamefont {M.}~\bibnamefont
  {Vandevraye}}, \bibinfo {author} {\bibfnamefont {P.}~\bibnamefont
  {Babilotte}}, \bibinfo {author} {\bibfnamefont {C.}~\bibnamefont {Drag}},\
  and\ \bibinfo {author} {\bibfnamefont {C.}~\bibnamefont {Blondel}},\
  }\bibfield  {title} {\bibinfo {title} {Laser measurement of the
  photodetachment cross section of {{${\mathrm{H}}^{\ensuremath{-}}$}} at the
  wavelength 1064 nm},\ }\href {https://doi.org/10.1103/PhysRevA.90.013411}
  {\bibfield  {journal} {\bibinfo  {journal} {Physical Review A}\ }\textbf
  {\bibinfo {volume} {90}},\ \bibinfo {pages} {013411} (\bibinfo {year}
  {2014})}\BibitemShut {NoStop}%
\bibitem [{\citenamefont {Bresteau}\ \emph {et~al.}(2017)\citenamefont
  {Bresteau}, \citenamefont {Blondel},\ and\ \citenamefont
  {Drag}}]{Bresteau2017}%
  \BibitemOpen
  \bibfield  {author} {\bibinfo {author} {\bibfnamefont {D.}~\bibnamefont
  {Bresteau}}, \bibinfo {author} {\bibfnamefont {C.}~\bibnamefont {Blondel}},\
  and\ \bibinfo {author} {\bibfnamefont {C.}~\bibnamefont {Drag}},\ }\bibfield
  {title} {\bibinfo {title} {Saturation of the photoneutralization of a
  {{$\mathrm{H}^-$}} beam in continuous operation},\ }\href
  {https://doi.org/10.1063/1.4995390} {\bibfield  {journal} {\bibinfo
  {journal} {Review of Scientific Instruments}\ }\textbf {\bibinfo {volume}
  {88}},\ \bibinfo {pages} {113103} (\bibinfo {year} {2017})}\BibitemShut
  {NoStop}%
\bibitem [{\citenamefont {Radics}\ \emph {et~al.}(2019)\citenamefont {Radics},
  \citenamefont {Janka}, \citenamefont {Cooke}, \citenamefont {Procureur},\
  and\ \citenamefont {Crivelli}}]{Radics2019}%
  \BibitemOpen
  \bibfield  {author} {\bibinfo {author} {\bibfnamefont {B.}~\bibnamefont
  {Radics}}, \bibinfo {author} {\bibfnamefont {G.}~\bibnamefont {Janka}},
  \bibinfo {author} {\bibfnamefont {D.~A.}\ \bibnamefont {Cooke}}, \bibinfo
  {author} {\bibfnamefont {S.}~\bibnamefont {Procureur}},\ and\ \bibinfo
  {author} {\bibfnamefont {P.}~\bibnamefont {Crivelli}},\ }\bibfield  {title}
  {\bibinfo {title} {Double hit reconstruction in large area multiplexed
  detectors},\ }\href {https://doi.org/10.1063/1.5109315} {\bibfield  {journal}
  {\bibinfo  {journal} {Review of Scientific Instruments}\ }\textbf {\bibinfo
  {volume} {90}},\ \bibinfo {pages} {093305} (\bibinfo {year}
  {2019})}\BibitemShut {NoStop}%
\bibitem [{\citenamefont {Rousselle}\ \emph
  {et~al.}(2022{\natexlab{a}})\citenamefont {Rousselle}, \citenamefont
  {Clad{\'{e}}}, \citenamefont {Guellati-Khelifa}, \citenamefont
  {Gu{\'{e}}rout},\ and\ \citenamefont {Reynaud}}]{Rousselle2022NJP}%
  \BibitemOpen
  \bibfield  {author} {\bibinfo {author} {\bibfnamefont {O.}~\bibnamefont
  {Rousselle}}, \bibinfo {author} {\bibfnamefont {P.}~\bibnamefont
  {Clad{\'{e}}}}, \bibinfo {author} {\bibfnamefont {S.}~\bibnamefont
  {Guellati-Khelifa}}, \bibinfo {author} {\bibfnamefont {R.}~\bibnamefont
  {Gu{\'{e}}rout}},\ and\ \bibinfo {author} {\bibfnamefont {S.}~\bibnamefont
  {Reynaud}},\ }\bibfield  {title} {\bibinfo {title} {Analysis of the timing of
  freely falling antihydrogen},\ }\href
  {https://doi.org/10.1088/1367-2630/ac5b57} {\bibfield  {journal} {\bibinfo
  {journal} {New Journal of Physics}\ }\textbf {\bibinfo {volume} {24}},\
  \bibinfo {pages} {033045} (\bibinfo {year} {2022}{\natexlab{a}})}\BibitemShut
  {NoStop}%
\bibitem [{\citenamefont {Rousselle}\ \emph
  {et~al.}(2022{\natexlab{b}})\citenamefont {Rousselle}, \citenamefont
  {Clad\'e}, \citenamefont {Guellati-Khelifa}, \citenamefont {Gu\'erout},\ and\
  \citenamefont {Reynaud}}]{Rousselle2022PRA}%
  \BibitemOpen
  \bibfield  {author} {\bibinfo {author} {\bibfnamefont {O.}~\bibnamefont
  {Rousselle}}, \bibinfo {author} {\bibfnamefont {P.}~\bibnamefont {Clad\'e}},
  \bibinfo {author} {\bibfnamefont {S.}~\bibnamefont {Guellati-Khelifa}},
  \bibinfo {author} {\bibfnamefont {R.}~\bibnamefont {Gu\'erout}},\ and\
  \bibinfo {author} {\bibfnamefont {S.}~\bibnamefont {Reynaud}},\ }\bibfield
  {title} {\bibinfo {title} {Improving the statistical analysis of
  anti-hydrogen free fall by using near edge events},\ }\href
  {https://doi.org/10.1103/PhysRevA.105.022821} {\bibfield  {journal} {\bibinfo
   {journal} {Physical Review A}\ }\textbf {\bibinfo {volume} {105}},\ \bibinfo
  {pages} {022821} (\bibinfo {year} {2022}{\natexlab{b}})}\BibitemShut
  {NoStop}%
\bibitem [{\citenamefont {Kasevich}\ and\ \citenamefont
  {Chu}(1991)}]{Kasevich1991}%
  \BibitemOpen
  \bibfield  {author} {\bibinfo {author} {\bibfnamefont {M.}~\bibnamefont
  {Kasevich}}\ and\ \bibinfo {author} {\bibfnamefont {S.}~\bibnamefont {Chu}},\
  }\bibfield  {title} {\bibinfo {title} {Atomic interferometry using stimulated
  {Raman} transitions},\ }\href {https://doi.org/10.1103/PhysRevLett.67.181}
  {\bibfield  {journal} {\bibinfo  {journal} {Physical Review Letters}\
  }\textbf {\bibinfo {volume} {67}},\ \bibinfo {pages} {181} (\bibinfo {year}
  {1991})}\BibitemShut {NoStop}%
\bibitem [{\citenamefont {Bord{\'e}}(2002)}]{Borde2002}%
  \BibitemOpen
  \bibfield  {author} {\bibinfo {author} {\bibfnamefont {C.~J.}\ \bibnamefont
  {Bord{\'e}}},\ }\bibfield  {title} {\bibinfo {title} {Atomic clocks and
  inertial sensors},\ }\href {https://doi.org/10.1088/0026-1394/39/5/5}
  {\bibfield  {journal} {\bibinfo  {journal} {Metrologia}\ }\textbf {\bibinfo
  {volume} {39}},\ \bibinfo {pages} {435} (\bibinfo {year} {2002})}\BibitemShut
  {NoStop}%
\bibitem [{\citenamefont {Merlet}\ \emph {et~al.}(2010)\citenamefont {Merlet},
  \citenamefont {Bodart}, \citenamefont {Malossi}, \citenamefont {Landragin},
  \citenamefont {Pereira Dos~Santos}, \citenamefont {Gitlein},\ and\
  \citenamefont {Timmen}}]{Merlet2010}%
  \BibitemOpen
  \bibfield  {author} {\bibinfo {author} {\bibfnamefont {S.}~\bibnamefont
  {Merlet}}, \bibinfo {author} {\bibfnamefont {Q.}~\bibnamefont {Bodart}},
  \bibinfo {author} {\bibfnamefont {N.}~\bibnamefont {Malossi}}, \bibinfo
  {author} {\bibfnamefont {A.}~\bibnamefont {Landragin}}, \bibinfo {author}
  {\bibfnamefont {F.}~\bibnamefont {Pereira Dos~Santos}}, \bibinfo {author}
  {\bibfnamefont {O.}~\bibnamefont {Gitlein}},\ and\ \bibinfo {author}
  {\bibfnamefont {L.}~\bibnamefont {Timmen}},\ }\bibfield  {title} {\bibinfo
  {title} {Comparison between two mobile absolute gravimeters: optical versus
  atomic interferometers},\ }\href {https://doi.org/10.1088/0026-1394/47/4/l01}
  {\bibfield  {journal} {\bibinfo  {journal} {Metrologia}\ }\textbf {\bibinfo
  {volume} {47}},\ \bibinfo {pages} {L9} (\bibinfo {year} {2010})}\BibitemShut
  {NoStop}%
\bibitem [{\citenamefont {Asenbaum}\ \emph {et~al.}(2020)\citenamefont
  {Asenbaum}, \citenamefont {Overstreet}, \citenamefont {Kim}, \citenamefont
  {Curti},\ and\ \citenamefont {Kasevich}}]{Asenbaum2020}%
  \BibitemOpen
  \bibfield  {author} {\bibinfo {author} {\bibfnamefont {P.}~\bibnamefont
  {Asenbaum}}, \bibinfo {author} {\bibfnamefont {C.}~\bibnamefont
  {Overstreet}}, \bibinfo {author} {\bibfnamefont {M.}~\bibnamefont {Kim}},
  \bibinfo {author} {\bibfnamefont {J.}~\bibnamefont {Curti}},\ and\ \bibinfo
  {author} {\bibfnamefont {M.}~\bibnamefont {Kasevich}},\ }\bibfield  {title}
  {\bibinfo {title} {Atom-interferometric test of the equivalence principle at
  the ${10}^{\ensuremath{-}12}$ level},\ }\href
  {https://doi.org/10.1103/PhysRevLett.125.191101} {\bibfield  {journal}
  {\bibinfo  {journal} {Physical Review Letters}\ }\textbf {\bibinfo {volume}
  {125}},\ \bibinfo {pages} {191101} (\bibinfo {year} {2020})}\BibitemShut
  {NoStop}%
\bibitem [{\citenamefont {Cr{\'e}pin}\ \emph
  {et~al.}(2019{\natexlab{a}})\citenamefont {Cr{\'e}pin}, \citenamefont
  {Christen}, \citenamefont {Gu{\'e}rout}, \citenamefont {Nesvizhevsky},
  \citenamefont {Voronin},\ and\ \citenamefont {Reynaud}}]{Crepin2019}%
  \BibitemOpen
  \bibfield  {author} {\bibinfo {author} {\bibfnamefont {P.-P.}\ \bibnamefont
  {Cr{\'e}pin}}, \bibinfo {author} {\bibfnamefont {C.}~\bibnamefont
  {Christen}}, \bibinfo {author} {\bibfnamefont {R.}~\bibnamefont
  {Gu{\'e}rout}}, \bibinfo {author} {\bibfnamefont {V.}~\bibnamefont
  {Nesvizhevsky}}, \bibinfo {author} {\bibfnamefont {A.}~\bibnamefont
  {Voronin}},\ and\ \bibinfo {author} {\bibfnamefont {S.}~\bibnamefont
  {Reynaud}},\ }\bibfield  {title} {\bibinfo {title} {Quantum interference test
  of the equivalence principle on antihydrogen},\ }\href
  {https://doi.org/10.1103/PhysRevA.99.042119} {\bibfield  {journal} {\bibinfo
  {journal} {Physical Review A}\ }\textbf {\bibinfo {volume} {99}},\ \bibinfo
  {pages} {042119} (\bibinfo {year} {2019}{\natexlab{a}})}\BibitemShut
  {NoStop}%
\bibitem [{\citenamefont {Nesvizhevsky}\ \emph {et~al.}(2009)\citenamefont
  {Nesvizhevsky}, \citenamefont {Voronin}, \citenamefont {Cubitt},\ and\
  \citenamefont {Protasov}}]{Nesvizhevsky2009}%
  \BibitemOpen
  \bibfield  {author} {\bibinfo {author} {\bibfnamefont {V.}~\bibnamefont
  {Nesvizhevsky}}, \bibinfo {author} {\bibfnamefont {A.}~\bibnamefont
  {Voronin}}, \bibinfo {author} {\bibfnamefont {R.}~\bibnamefont {Cubitt}},\
  and\ \bibinfo {author} {\bibfnamefont {K.}~\bibnamefont {Protasov}},\
  }\bibfield  {title} {\bibinfo {title} {Neutron whispering gallery},\ }\href
  {https://doi.org/10.1038/nphys1478} {\bibfield  {journal} {\bibinfo
  {journal} {Nature Physics}\ }\textbf {\bibinfo {volume} {6}},\ \bibinfo
  {pages} {114} (\bibinfo {year} {2009})}\BibitemShut {NoStop}%
\bibitem [{\citenamefont {Nesvizhevsky}\ \emph {et~al.}(2002)\citenamefont
  {Nesvizhevsky}, \citenamefont {B{\"o}rner}, \citenamefont {Petukhov},
  \citenamefont {Abele}, \citenamefont {Bae{\ss}ler}, \citenamefont {Rue{\ss}},
  \citenamefont {St{\"o}ferle}, \citenamefont {Westphal}, \citenamefont
  {Gagarski}, \citenamefont {Petrov},\ and\ \citenamefont
  {Strelkov}}]{Nesvizhevsky2002}%
  \BibitemOpen
  \bibfield  {author} {\bibinfo {author} {\bibfnamefont {V.}~\bibnamefont
  {Nesvizhevsky}}, \bibinfo {author} {\bibfnamefont {H.}~\bibnamefont
  {B{\"o}rner}}, \bibinfo {author} {\bibfnamefont {A.}~\bibnamefont
  {Petukhov}}, \bibinfo {author} {\bibfnamefont {H.}~\bibnamefont {Abele}},
  \bibinfo {author} {\bibfnamefont {S.}~\bibnamefont {Bae{\ss}ler}}, \bibinfo
  {author} {\bibfnamefont {F.}~\bibnamefont {Rue{\ss}}}, \bibinfo {author}
  {\bibfnamefont {T.}~\bibnamefont {St{\"o}ferle}}, \bibinfo {author}
  {\bibfnamefont {A.}~\bibnamefont {Westphal}}, \bibinfo {author}
  {\bibfnamefont {A.}~\bibnamefont {Gagarski}}, \bibinfo {author}
  {\bibfnamefont {G.}~\bibnamefont {Petrov}},\ and\ \bibinfo {author}
  {\bibfnamefont {A.}~\bibnamefont {Strelkov}},\ }\bibfield  {title} {\bibinfo
  {title} {Quantum states of neutrons in the {{Earth}}'s gravitational field},\
  }\href {https://doi.org/10.1038/415297a} {\bibfield  {journal} {\bibinfo
  {journal} {Nature}\ }\textbf {\bibinfo {volume} {415}},\ \bibinfo {pages}
  {297} (\bibinfo {year} {2002})}\BibitemShut {NoStop}%
\bibitem [{\citenamefont {Nesvizhevsky}\ \emph {et~al.}(2003)\citenamefont
  {Nesvizhevsky}, \citenamefont {B{\"o}rner}, \citenamefont {Gagarski},
  \citenamefont {Petoukhov}, \citenamefont {Petrov}, \citenamefont {Abele},
  \citenamefont {Bae{\ss}ler}, \citenamefont {Divkovic}, \citenamefont
  {Rue{\ss}}, \citenamefont {St{\"o}ferle}, \citenamefont {Westphal},
  \citenamefont {Strelkov}, \citenamefont {Protasov},\ and\ \citenamefont
  {Voronin}}]{Nesvizhevsky2003}%
  \BibitemOpen
  \bibfield  {author} {\bibinfo {author} {\bibfnamefont {V.}~\bibnamefont
  {Nesvizhevsky}}, \bibinfo {author} {\bibfnamefont {H.}~\bibnamefont
  {B{\"o}rner}}, \bibinfo {author} {\bibfnamefont {A.}~\bibnamefont
  {Gagarski}}, \bibinfo {author} {\bibfnamefont {A.}~\bibnamefont {Petoukhov}},
  \bibinfo {author} {\bibfnamefont {G.}~\bibnamefont {Petrov}}, \bibinfo
  {author} {\bibfnamefont {H.}~\bibnamefont {Abele}}, \bibinfo {author}
  {\bibfnamefont {S.}~\bibnamefont {Bae{\ss}ler}}, \bibinfo {author}
  {\bibfnamefont {G.}~\bibnamefont {Divkovic}}, \bibinfo {author}
  {\bibfnamefont {F.}~\bibnamefont {Rue{\ss}}}, \bibinfo {author}
  {\bibfnamefont {T.}~\bibnamefont {St{\"o}ferle}}, \bibinfo {author}
  {\bibfnamefont {A.}~\bibnamefont {Westphal}}, \bibinfo {author}
  {\bibfnamefont {A.}~\bibnamefont {Strelkov}}, \bibinfo {author}
  {\bibfnamefont {K.}~\bibnamefont {Protasov}},\ and\ \bibinfo {author}
  {\bibfnamefont {A.}~\bibnamefont {Voronin}},\ }\bibfield  {title} {\bibinfo
  {title} {Measurement of quantum states of neutrons in the {Earth}'s
  gravitational field},\ }\href {https://doi.org/10.1103/PhysRevD.67.102002}
  {\bibfield  {journal} {\bibinfo  {journal} {Physical Review D}\ }\textbf
  {\bibinfo {volume} {67}},\ \bibinfo {pages} {102002} (\bibinfo {year}
  {2003})}\BibitemShut {NoStop}%
\bibitem [{\citenamefont {Nesvizhevsky}\ \emph {et~al.}(2005)\citenamefont
  {Nesvizhevsky}, \citenamefont {Petukhov}, \citenamefont {Borner},
  \citenamefont {Baranova}, \citenamefont {Gagarski}, \citenamefont {Petrov},
  \citenamefont {Protasov}, \citenamefont {Voronin}, \citenamefont {Baessler},
  \citenamefont {Abele}, \citenamefont {Westphal},\ and\ \citenamefont
  {Lucovac}}]{Nesvizhevsky2005}%
  \BibitemOpen
  \bibfield  {author} {\bibinfo {author} {\bibfnamefont {V.}~\bibnamefont
  {Nesvizhevsky}}, \bibinfo {author} {\bibfnamefont {A.}~\bibnamefont
  {Petukhov}}, \bibinfo {author} {\bibfnamefont {H.}~\bibnamefont {Borner}},
  \bibinfo {author} {\bibfnamefont {T.}~\bibnamefont {Baranova}}, \bibinfo
  {author} {\bibfnamefont {A.}~\bibnamefont {Gagarski}}, \bibinfo {author}
  {\bibfnamefont {G.}~\bibnamefont {Petrov}}, \bibinfo {author} {\bibfnamefont
  {K.}~\bibnamefont {Protasov}}, \bibinfo {author} {\bibfnamefont
  {A.}~\bibnamefont {Voronin}}, \bibinfo {author} {\bibfnamefont
  {S.}~\bibnamefont {Baessler}}, \bibinfo {author} {\bibfnamefont
  {H.}~\bibnamefont {Abele}}, \bibinfo {author} {\bibfnamefont
  {A.}~\bibnamefont {Westphal}},\ and\ \bibinfo {author} {\bibfnamefont
  {L.}~\bibnamefont {Lucovac}},\ }\bibfield  {title} {\bibinfo {title} {Study
  of the neutron quantum states in the gravity field},\ }\href
  {https://doi.org/10.1140/epjc/s2005-02135-y} {\bibfield  {journal} {\bibinfo
  {journal} {European Physical Journal C}\ }\textbf {\bibinfo {volume} {40}},\
  \bibinfo {pages} {479} (\bibinfo {year} {2005})}\BibitemShut {NoStop}%
\bibitem [{\citenamefont {Froelich}\ and\ \citenamefont
  {Voronin}(2012)}]{Froelich2012}%
  \BibitemOpen
  \bibfield  {author} {\bibinfo {author} {\bibfnamefont {P.}~\bibnamefont
  {Froelich}}\ and\ \bibinfo {author} {\bibfnamefont {A.}~\bibnamefont
  {Voronin}},\ }\bibfield  {title} {\bibinfo {title} {Interaction of
  antihydrogen with ordinary atoms and solid surfaces},\ }\href
  {https://doi.org/10.1007/s10751-012-0627-2} {\bibfield  {journal} {\bibinfo
  {journal} {Hyperfine Interactions}\ }\textbf {\bibinfo {volume} {213}},\
  \bibinfo {pages} {115–127} (\bibinfo {year} {2012})}\BibitemShut {NoStop}%
\bibitem [{\citenamefont {Dufour}\ \emph {et~al.}(2013)\citenamefont {Dufour},
  \citenamefont {G\'erardin}, \citenamefont {Gu\'erout}, \citenamefont
  {Lambrecht}, \citenamefont {Nesvizhevsky}, \citenamefont {Reynaud},\ and\
  \citenamefont {Voronin}}]{Dufour2013}%
  \BibitemOpen
  \bibfield  {author} {\bibinfo {author} {\bibfnamefont {G.}~\bibnamefont
  {Dufour}}, \bibinfo {author} {\bibfnamefont {A.}~\bibnamefont {G\'erardin}},
  \bibinfo {author} {\bibfnamefont {R.}~\bibnamefont {Gu\'erout}}, \bibinfo
  {author} {\bibfnamefont {A.}~\bibnamefont {Lambrecht}}, \bibinfo {author}
  {\bibfnamefont {V.~V.}\ \bibnamefont {Nesvizhevsky}}, \bibinfo {author}
  {\bibfnamefont {S.}~\bibnamefont {Reynaud}},\ and\ \bibinfo {author}
  {\bibfnamefont {A.~Y.}\ \bibnamefont {Voronin}},\ }\bibfield  {title}
  {\bibinfo {title} {Quantum reflection of antihydrogen from the {Casimir}
  potential above matter slabs},\ }\href
  {https://doi.org/10.1103/PhysRevA.87.012901} {\bibfield  {journal} {\bibinfo
  {journal} {Physical Review A}\ }\textbf {\bibinfo {volume} {87}},\ \bibinfo
  {pages} {012901} (\bibinfo {year} {2013})}\BibitemShut {NoStop}%
\bibitem [{\citenamefont {Jurisch}\ and\ \citenamefont
  {Friedrich}(2006)}]{Jurisch2006}%
  \BibitemOpen
  \bibfield  {author} {\bibinfo {author} {\bibfnamefont {A.}~\bibnamefont
  {Jurisch}}\ and\ \bibinfo {author} {\bibfnamefont {H.}~\bibnamefont
  {Friedrich}},\ }\bibfield  {title} {\bibinfo {title} {Realistic model for a
  quantum reflection trap},\ }\href
  {https://doi.org/10.1016/j.physleta.2005.09.014} {\bibfield  {journal}
  {\bibinfo  {journal} {Physics Letters A}\ }\textbf {\bibinfo {volume}
  {349}},\ \bibinfo {pages} {230} (\bibinfo {year} {2006})}\BibitemShut
  {NoStop}%
\bibitem [{\citenamefont {Madronero}\ and\ \citenamefont
  {Friedrich}(2007)}]{Madronero2007}%
  \BibitemOpen
  \bibfield  {author} {\bibinfo {author} {\bibfnamefont {J.}~\bibnamefont
  {Madronero}}\ and\ \bibinfo {author} {\bibfnamefont {H.}~\bibnamefont
  {Friedrich}},\ }\bibfield  {title} {\bibinfo {title} {Influence of realistic
  atom wall potentials in quantum reflection traps},\ }\href
  {https://doi.org/10.1103/PhysRevA.75.022902} {\bibfield  {journal} {\bibinfo
  {journal} {Physical Review A}\ }\textbf {\bibinfo {volume} {75}},\ \bibinfo
  {pages} {022902} (\bibinfo {year} {2007})}\BibitemShut {NoStop}%
\bibitem [{\citenamefont {Nesvizhevsky}\ and\ \citenamefont
  {Voronin}(2015)}]{Nesvizhevsky2015}%
  \BibitemOpen
  \bibfield  {author} {\bibinfo {author} {\bibfnamefont {V.}~\bibnamefont
  {Nesvizhevsky}}\ and\ \bibinfo {author} {\bibfnamefont {A.~Y.}\ \bibnamefont
  {Voronin}},\ }\href@noop {} {\emph {\bibinfo {title} {Surprising quantum
  bounces}}}\ (\bibinfo  {publisher} {Imperial College Press},\ \bibinfo {year}
  {2015})\BibitemShut {NoStop}%
\bibitem [{\citenamefont {Olver}\ \emph {et~al.}(2010)\citenamefont {Olver},
  \citenamefont {Lozier}, \citenamefont {Boisvert},\ and\ \citenamefont
  {Clark}}]{Olver2010}%
  \BibitemOpen
  \bibfield  {author} {\bibinfo {author} {\bibfnamefont {F.~W.~J.}\
  \bibnamefont {Olver}}, \bibinfo {author} {\bibfnamefont {D.~W.}\ \bibnamefont
  {Lozier}}, \bibinfo {author} {\bibfnamefont {R.~F.}\ \bibnamefont
  {Boisvert}},\ and\ \bibinfo {author} {\bibfnamefont {C.~W.}\ \bibnamefont
  {Clark}},\ }\bibfield  {title} {\bibinfo {title} {{NIST} handbook of
  mathematical functions}\ }(\bibinfo  {publisher} {NIST},\ \bibinfo {year}
  {2010})\BibitemShut {NoStop}%
\bibitem [{\citenamefont {Vall{\'e}e}\ and\ \citenamefont
  {Soares}(2004)}]{Vallee2004}%
  \BibitemOpen
  \bibfield  {author} {\bibinfo {author} {\bibfnamefont {O.}~\bibnamefont
  {Vall{\'e}e}}\ and\ \bibinfo {author} {\bibfnamefont {M.}~\bibnamefont
  {Soares}},\ }\href@noop {} {\emph {\bibinfo {title} {Airy functions and
  applications to physics}}}\ (\bibinfo  {publisher} {Imperial College Press},\
  \bibinfo {year} {2004})\BibitemShut {NoStop}%
\bibitem [{\citenamefont {Meyerovich}\ and\ \citenamefont
  {Nesvizhevsky}(2006)}]{Meyerovich2006}%
  \BibitemOpen
  \bibfield  {author} {\bibinfo {author} {\bibfnamefont {A.~E.}\ \bibnamefont
  {Meyerovich}}\ and\ \bibinfo {author} {\bibfnamefont {V.~V.}\ \bibnamefont
  {Nesvizhevsky}},\ }\bibfield  {title} {\bibinfo {title} {Gravitational
  quantum states of neutrons in a rough waveguide},\ }\href
  {https://doi.org/10.1103/PhysRevA.73.063616} {\bibfield  {journal} {\bibinfo
  {journal} {Physical Review A}\ }\textbf {\bibinfo {volume} {73}},\ \bibinfo
  {pages} {063616} (\bibinfo {year} {2006})}\BibitemShut {NoStop}%
\bibitem [{\citenamefont {Escobar}\ \emph {et~al.}(2014)\citenamefont
  {Escobar}, \citenamefont {Lamy}, \citenamefont {Meyerovich},\ and\
  \citenamefont {Nesvizhevsky}}]{Escobar2014}%
  \BibitemOpen
  \bibfield  {author} {\bibinfo {author} {\bibfnamefont {M.}~\bibnamefont
  {Escobar}}, \bibinfo {author} {\bibfnamefont {F.}~\bibnamefont {Lamy}},
  \bibinfo {author} {\bibfnamefont {A.~E.}\ \bibnamefont {Meyerovich}},\ and\
  \bibinfo {author} {\bibfnamefont {V.~V.}\ \bibnamefont {Nesvizhevsky}},\
  }\bibfield  {title} {\bibinfo {title} {Rough mirror as a quantum state
  selector: Analysis and design},\ }\href {https://doi.org/10.1155/2014/764182}
  {\bibfield  {journal} {\bibinfo  {journal} {Advances in High Energy Physics}\
  }\textbf {\bibinfo {volume} {2014}},\ \bibinfo {pages} {764182} (\bibinfo
  {year} {2014})}\BibitemShut {NoStop}%
\bibitem [{\citenamefont {Dufour}\ \emph {et~al.}(2014)\citenamefont {Dufour},
  \citenamefont {Debu}, \citenamefont {Lambrecht}, \citenamefont
  {Nesvizhevsky}, \citenamefont {Reynaud},\ and\ \citenamefont
  {Voronin}}]{Dufour2014}%
  \BibitemOpen
  \bibfield  {author} {\bibinfo {author} {\bibfnamefont {G.}~\bibnamefont
  {Dufour}}, \bibinfo {author} {\bibfnamefont {P.}~\bibnamefont {Debu}},
  \bibinfo {author} {\bibfnamefont {A.}~\bibnamefont {Lambrecht}}, \bibinfo
  {author} {\bibfnamefont {V.~V.}\ \bibnamefont {Nesvizhevsky}}, \bibinfo
  {author} {\bibfnamefont {S.}~\bibnamefont {Reynaud}},\ and\ \bibinfo {author}
  {\bibfnamefont {A.~Y.}\ \bibnamefont {Voronin}},\ }\bibfield  {title}
  {\bibinfo {title} {Shaping the distribution of vertical velocities of
  antihydrogen in {GBAR}},\ }\href
  {https://doi.org/10.1140/epjc/s10052-014-2731-8} {\bibfield  {journal}
  {\bibinfo  {journal} {The European Physical Journal C}\ }\textbf {\bibinfo
  {volume} {74}},\ \bibinfo {pages} {2731} (\bibinfo {year}
  {2014})}\BibitemShut {NoStop}%
\bibitem [{\citenamefont {Storey}\ and\ \citenamefont
  {Cohen-Tannoudji}(1994)}]{Storey1994}%
  \BibitemOpen
  \bibfield  {author} {\bibinfo {author} {\bibfnamefont {P.}~\bibnamefont
  {Storey}}\ and\ \bibinfo {author} {\bibfnamefont {C.}~\bibnamefont
  {Cohen-Tannoudji}},\ }\bibfield  {title} {\bibinfo {title} {The {Feynman}
  path-integral approach to atomic interferometry - a tutorial},\ }\href
  {https://doi.org/10.1051/jp2:1994103} {\bibfield  {journal} {\bibinfo
  {journal} {Journal de Physique II}\ }\textbf {\bibinfo {volume} {4}},\
  \bibinfo {pages} {1999} (\bibinfo {year} {1994})}\BibitemShut {NoStop}%
\bibitem [{\citenamefont {Fr\'echet}(1943)}]{Frechet1943}%
  \BibitemOpen
  \bibfield  {author} {\bibinfo {author} {\bibfnamefont {M.}~\bibnamefont
  {Fr\'echet}},\ }\bibfield  {title} {\bibinfo {title} {Sur l'extension de
  certaines \'evaluations statistiques au cas de petits \'echantillons},\
  }\href {https://doi.org/10.2307/1401114} {\bibfield  {journal} {\bibinfo
  {journal} {Review of the International Statistical Institute}\ }\textbf
  {\bibinfo {volume} {11}},\ \bibinfo {pages} {182} (\bibinfo {year}
  {1943})}\BibitemShut {NoStop}%
\bibitem [{\citenamefont {Cram\'er}(1999)}]{Cramer1999}%
  \BibitemOpen
  \bibfield  {author} {\bibinfo {author} {\bibfnamefont {H.}~\bibnamefont
  {Cram\'er}},\ }\href@noop {} {\emph {\bibinfo {title} {Mathematical Methods
  of Statistics (new edition)}}}\ (\bibinfo  {publisher} {{Princeton University
  Press}},\ \bibinfo {year} {1999})\BibitemShut {NoStop}%
\bibitem [{\citenamefont {R{\'e}fr{\'e}gier}(2004)}]{Refregier2004}%
  \BibitemOpen
  \bibfield  {author} {\bibinfo {author} {\bibfnamefont {P.}~\bibnamefont
  {R{\'e}fr{\'e}gier}},\ }\href@noop {} {\emph {\bibinfo {title} {Noise Theory
  and Application to Physics: From Fluctuations to Information}}},\ Advanced
  Texts in Physics\ (\bibinfo  {publisher} {Springer},\ \bibinfo {address} {New
  York},\ \bibinfo {year} {2004})\BibitemShut {NoStop}%
\bibitem [{\citenamefont {Cr{\'e}pin}\ \emph {et~al.}(2017)\citenamefont
  {Cr{\'e}pin}, \citenamefont {Dufour}, \citenamefont {Gu{\'e}rout},
  \citenamefont {Lambrecht},\ and\ \citenamefont {Reynaud}}]{Crepin2017}%
  \BibitemOpen
  \bibfield  {author} {\bibinfo {author} {\bibfnamefont {P.-P.}\ \bibnamefont
  {Cr{\'e}pin}}, \bibinfo {author} {\bibfnamefont {G.}~\bibnamefont {Dufour}},
  \bibinfo {author} {\bibfnamefont {R.}~\bibnamefont {Gu{\'e}rout}}, \bibinfo
  {author} {\bibfnamefont {A.}~\bibnamefont {Lambrecht}},\ and\ \bibinfo
  {author} {\bibfnamefont {S.}~\bibnamefont {Reynaud}},\ }\bibfield  {title}
  {\bibinfo {title} {{Casimir-Polder} shifts on quantum levitation states},\
  }\href {https://doi.org/10.1103/PhysRevA.95.032501} {\bibfield  {journal}
  {\bibinfo  {journal} {Physical Review A}\ }\textbf {\bibinfo {volume} {95}},\
  \bibinfo {pages} {032501} (\bibinfo {year} {2017})}\BibitemShut {NoStop}%
\bibitem [{\citenamefont {Cr{\'e}pin}\ \emph
  {et~al.}(2019{\natexlab{b}})\citenamefont {Cr{\'e}pin}, \citenamefont
  {Gu{\'e}rout},\ and\ \citenamefont {Reynaud}}]{Crepin2019d}%
  \BibitemOpen
  \bibfield  {author} {\bibinfo {author} {\bibfnamefont {P.-P.}\ \bibnamefont
  {Cr{\'e}pin}}, \bibinfo {author} {\bibfnamefont {R.}~\bibnamefont
  {Gu{\'e}rout}},\ and\ \bibinfo {author} {\bibfnamefont {S.}~\bibnamefont
  {Reynaud}},\ }\bibfield  {title} {\bibinfo {title} {Improved effective range
  expansion for {Casimir-Polder} potential},\ }\href
  {https://doi.org/10.1140/epjd/e2019-100196-2} {\bibfield  {journal} {\bibinfo
   {journal} {The European Physical Journal D}\ }\textbf {\bibinfo {volume}
  {73}},\ \bibinfo {pages} {256} (\bibinfo {year}
  {2019}{\natexlab{b}})}\BibitemShut {NoStop}%
\end{thebibliography}%

\end{document}